\documentclass{article}

\usepackage{microtype}
\usepackage{graphicx}
\usepackage{subcaption}
\usepackage{booktabs}
\usepackage{xurl}

\usepackage{hyperref}

\usepackage[accepted]{icml2026}

\usepackage{xcolor}
\usepackage[most]{tcolorbox}

\usepackage{amsmath}
\usepackage{amssymb}
\usepackage{mathtools}
\usepackage{amsthm}
\usepackage[capitalize,noabbrev]{cleveref}

\usepackage{enumitem}
\usepackage{listings}
\setlist[itemize]{leftmargin=1.1em}
\setlist[enumerate]{leftmargin=1.2em}

\definecolor{promptbg}{RGB}{249,249,247}
\definecolor{promptframe}{RGB}{205,205,205}
\newtcblisting{promptbox}{
  breakable,
  enhanced,
  colback=promptbg,
  colframe=promptframe,
  boxrule=0.45pt,
  arc=2pt,
  boxsep=0pt,
  left=0.7em,
  right=0.7em,
  top=0.55em,
  bottom=0.55em,
  before skip=0.8em,
  after skip=0.8em,
  pad at break*=1mm,
  listing only,
  listing options={
    basicstyle=\ttfamily\footnotesize,
    breaklines=true,
    breakatwhitespace=false,
    columns=fullflexible,
    keepspaces=true,
    showstringspaces=false
  }
}

\usepackage{tikz}
\usetikzlibrary{positioning,arrows.meta,calc,fit,backgrounds}

\theoremstyle{definition}

\newcommand{\Uniprot}{\textsc{UniProt}}
\newcommand{\IEDB}{\textsc{IEDB}}

\newcommand{\LLM}{\textsc{LLM}}
\newcommand{\pLDDT}{\textsc{pLDDT}}

\newcommand{\sysname}{Site4Drug}
\newcommand{\Sys}{\textbf{\sysname}}

\icmltitlerunning{\sysname: Predicting Drug-Binding Target Sites with an AI Agent}

\newif\ifshowfrontmatterextras
\showfrontmatterextrasfalse
\ifdefined\isaccepted\showfrontmatterextrastrue\fi
\ifdefined\ispreprint\showfrontmatterextrastrue\fi

\begin{document}

\twocolumn[
\icmltitle{\sysname: Predicting Drug-Binding Target Sites with an AI Agent}

\begin{icmlauthorlist}
\icmlauthor{Taehan~Kim\textsuperscript{\textdagger}}{ucb}
\icmlauthor{Sarrah Rose Mikhail Leung}{ucb}
\icmlauthor{Bharat~Mekala}{ucb}
\icmlauthor{Jeongbin~Park\textsuperscript{\textdagger}}{umich}
\end{icmlauthorlist}

\icmlaffiliation{ucb}{UC Berkeley}
\icmlaffiliation{umich}{University of Michigan}

\icmlcorrespondingauthor{Taehan~Kim}{terry.kim@berkeley.edu}
\icmlcorrespondingauthor{Jeongbin~Park}{jeongbp@umich.edu}

\icmlkeywords{protein targeting, epitope discovery, pocket discovery, auditability, agentic systems}
\vskip 0.3in
]

\ifshowfrontmatterextras
  \printAffiliationsAndNotice{\textsuperscript{\textdagger}Corresponding authors.}
\else
  \printAffiliationsAndNotice{}
\fi

\begin{abstract}
Selecting \emph{where} to intervene on a protein (i.e., choosing a \textbf{targetable site}) is often a more ambiguous and failure-prone bottleneck than selecting \emph{what} binds, especially for membrane proteins where accessibility, topology, and post-translational modifications (PTMs) constrain actionable regions. We present \Sys, a modality-aware site-finding agent that outputs a ranked list of \textbf{targetable regions} with explicit constraints, evidence summaries, risk flags, and a traceable decision log.

Rather than requiring users to specify the drug modality upfront, \sysname\ can \textbf{recommend a binding modality} (e.g., antibody/peptide-like vs small-molecule) from the same evidence used for site discovery, including topology, hydropathy, PTM propensity, disulfides, domain context, and sequence.
Importantly, this evidence is applied consistently across modalities, including small-molecule pocket discovery, to avoid selecting chemically plausible but biologically occluded sites.
\end{abstract}

\section{Introduction}
Many discovery pipelines assume a binding site is known and focus on docking, screening, or binder generation. For example, this includes models like BoltzGen \cite{stark2025boltzgen} for binder generation and CLIP-style pocket scoring models (e.g., DrugCLIP \cite{gao2023drugclip}, BindCLIP \cite{qiao2026bindclip}) that evaluate candidates given a predefined site.

In practice, teams often stall earlier: \emph{where and with what modality should we target on the protein?}
For membrane proteins this is especially difficult: only certain regions are physically accessible, topology predictions can disagree, and PTMs (notably glycosylation) can occlude or destabilize candidate epitopes.
When downstream screening fails, teams may not know whether the binding model failed or the \emph{site choice} was wrong, partly because the reasoning behind site selection is rarely logged or can be arbitrary.


Curated resources can help, but they are not absolute ground truths for therapeutic site selection.
For example, \IEDB\ \cite{vita2025immune} aggregates immune epitope evidence from highly specialized experimental settings (e.g., T cell and B cell assays) that are context-dependent and inherently incomplete; critically, these assays were not designed to exhaustively enumerate \emph{drug-actionable} binding sites on proteins. 

On the small-molecule side, ``known pockets'' are often operationally defined by taking residues contacting a co-crystallized ligand and padding by a fixed neighborhood.
Such definitions can be brittle: 
a limited number of known ligand-bound structures leaves the pocket undefined (no prior discoveries). This is particularly important for cancer neoantigens and mutated viral antigens.
Together, these limitations motivate 
epitope/pocket discovery that is grounded in feasibility constraints along with reasoning traces. 

Conventional approaches such as fpocket \cite{le2009fpocket} and RAPID-Net \cite{balytskyi2025rapid} rely on geometric features or learned structural representations, and thus can primarily be applied to the identification of canonical binding pockets. As a result, they are limited in their ability to incorporate diverse metadata or to define sites for alternative drug modalities. Recently, large language models (LLMs) have been increasingly utilized in tasks such as protein structure prediction and drug-genomic response prediction. Moreover, their ability to flexibly integrate diverse forms of unstructured data suggests that they should be considered for adoption in site definition tasks.

Additionally, LLM-based AI agentic systems are being introduced into AI-driven scientific discovery research (e.g., AI-Descartes \cite{cornelio2023ai_descartes}, Theorizer \cite{jansen2026literature}, AutoDiscovery \cite{agarwal2026autodiscovery}, The AI Scientist \cite{lu2026towards}, AI co-scientist \cite{gottweis2026cscientist}, The Virtual Lab \cite{swanson2024virtuallab}), because they require repeated execution of LLMs to achieve reproducible scientific conclusions. Therefore, we explore the potential of drug site detection as an LLM agentic system.

\begin{figure*}[!t]
  \centering
  \includegraphics[width=0.65\textwidth]{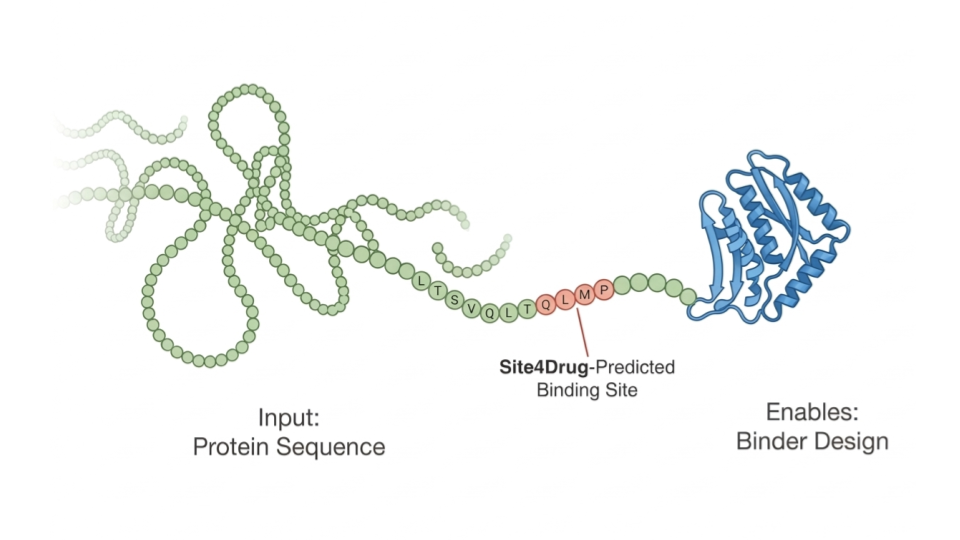}
  \caption{Site4Drug discovers a potential binding region for binder design.}
  \label{fig:cover_image}
\end{figure*}

Thus, \sysname\ attempts to provide a solution for the upstream bottleneck. It reframes site selection as a \textbf{constraint-first, evidence-integrated decision problem}.
Given a protein sequence, it proposes candidate regions, ranks them, and emits a structured report:
what constraints were applied, what evidence supported each candidate, and what risks remain.
This makes the site-selection step auditable and debuggable. With these results, third-party tools can also be invoked as part of an agentic system as summarized in Figure~\ref{fig:cover_image}.


\section{Problem Setup}
\subsection{Inputs and Outputs}
\textbf{Input.}
A target protein specified by an amino-acid sequence $x_{1:L}$.

\textbf{Output.}
\sysname\ returns a structured, auditable site-selection result consisting of:
\begin{itemize}
  \item \textbf{Recommended binding modality:}
  a categorical recommendation $\hat{m} \in \{\textsf{epitope}, \textsf{pocket}, \textsf{other}\}$ with an evidence-backed rationale and uncertainty notes.\footnote{\textsf{Epitope} refers to where antibodies or peptide binders attach, usually on extracellular protruding regions of membrane proteins and often tied to immune responses. \textsf{Pocket} refers to where small molecules bind, typically in intracellular proteins or membrane channels, covering immune as well as general signaling and metabolic pathways. \textsf{Other} covers cases where the evidence suggests non-standard or mixed modes (e.g., ambiguous accessibility or competing feasibility constraints).}

  \item \textbf{Ranked candidates with scores:}
  a list of $K$ candidates $\{r_k\}_{k=1}^K$ with per-candidate score $S(r_k)$:
  \begin{itemize}
    \item \textbf{Accessibility/topology label:} a coarse label derived from hydropathy and transmembrane (TM) overlap;
    \item \textbf{Evidence summary:} key signals supporting or penalizing the region (hydropathy, TM overlap, typed PTM masks, motif hits, cysteine/disulfide context);
    \item \textbf{Flags:} typed annotations including \emph{risk flags} (e.g., ``TM-overlap'', ``PTM-overlap'', ``glyco-mask-overlap'', ``disulfide-constrained'', ``motif-overlap'') 

  \end{itemize}
\end{itemize}
A representative output report is shown in the Appendix~\ref{app:reportss}, Fig.~\ref{fig:prediction_interface}.
\subsection{Method Overview: Two Modules}
\sysname\ potentially offers two modules:
\begin{enumerate}
  \item \textbf{Module 1 (Main Discovery Engine):} evidence extraction $\rightarrow$ constraint-first candidate generation $\rightarrow$ scoring/ranking with risk flags and audit log.
  \item \textbf{Module 2 (Potential Design Handoff):} route top candidates to modality-appropriate downstream design/scoring tools (e.g., peptide-binder generation in epitope mode; pocket-aware ligand scoring in pocket mode).
\end{enumerate}

\begin{figure*}[!t]
  \centering
  \includegraphics[width=0.50\textwidth]{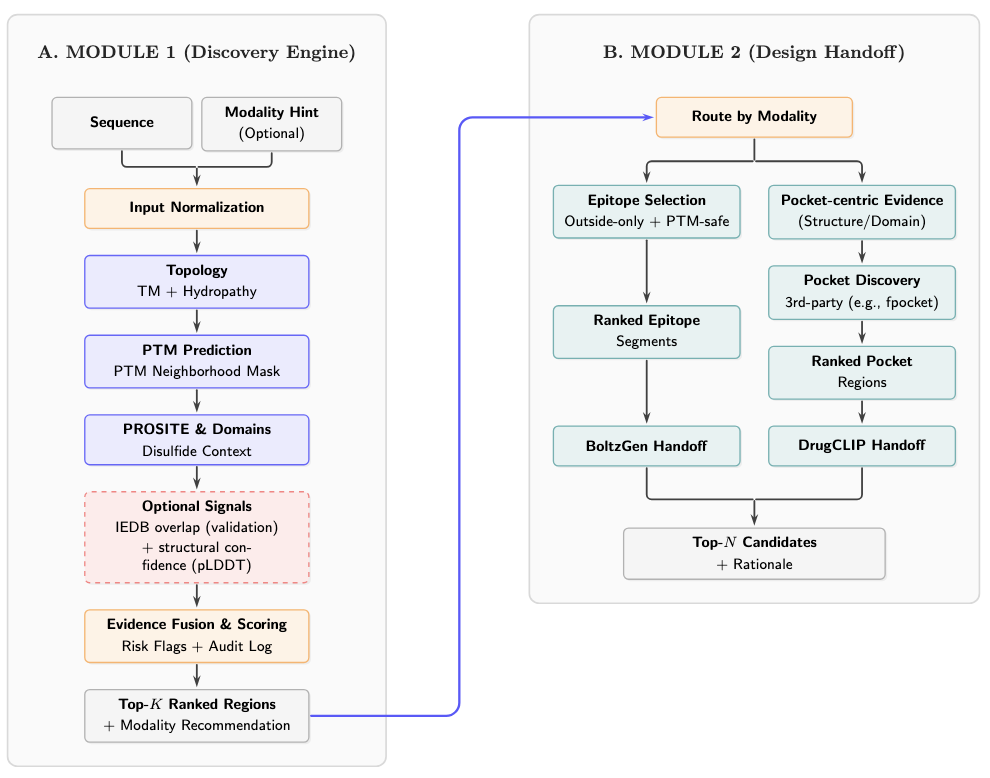}
  \caption{Two-module pipeline: constraint-first discovery (Module 1) followed by modality-specific design handoff (Module 2).}
  \label{fig:overview_pipeline}
\end{figure*}

\section{Module 1: Evidence Aggregation and Region Discovery}

As summarized in Figure~\ref{fig:overview_pipeline}, Module~1 constructs a residue- and region-level evidence summary from three signal classes: i) topology/accessibility priors, ii) PTM risk signals, and iii) motif/domain context. These signals are used to guide 1) binding site candidate proposal, 2) validate proposed spans, and 3) attach interpretable risk annotations.

\subsection{Evidence Extractors}
Given a protein sequence, \sysname\ computes three classes of feasibility-aware signals that are exposed to the proposal stage and retained for downstream validation.

\paragraph{I. Coarse topology + hydropathy.}
\sysname\  computes a sliding-window Kyte--Doolittle hydropathy profile \cite{kyte1982simple} together with a heuristic TM detector to derive a coarse accessibility prior from sequence alone. Regions overlapping detected TM segments are labeled \texttt{tmd}/\texttt{restricted}, whereas non-overlapping regions are labeled \texttt{outside}/\texttt{exposed}, with confidence determined by the hydropathy margin from the TM threshold.

\paragraph{II. PTM prediction + neighborhood masking.}
\sysname\  uses PTM calls as feasibility constraints of targetability. PTM sites are obtained from a MusiteDeep-backed extractor \cite{wang2020musitedeep} with rule-based augmentation, and each site is expanded into a typed local mask. We then record whether a candidate overlaps these masks, together with overlap counts by PTM type and local PTM density. For example, if MusiteDeep predicts a phosphoserine/phosphothreonine site at residue 211, we expand it into a typed local mask (e.g., residues 208--214) and record whether a candidate region overlaps that mask.

\paragraph{III. Motif hits + cysteine/disulfide proxy.}
\sysname\  also queries ScanProsite \cite{de2006scanprosite} and records motif spans as cautionary context for targetability. Overlap with a known motif yields a \texttt{motif-overlap} annotation, since such regions may be conserved or functionally constrained. We also track cysteine counts as a lightweight proxy for disulfide-constrained segments; regions containing multiple cysteines receive a \texttt{disulfide-constrained} flag.

\subsection{Candidate Generation and Ranking}
Conditioned on the sequence and the compact evidence summary described above, the language model proposes a ranked JSON list of candidate regions directly. After filtering invalid candidates, each is annotated with topology/accessibility evidence, PTM overlap, motif overlap, cysteine count, typed risk flags, and a heuristic score.

\subsection{Reranking with Specialist}
A specialist panel provides a secondary adjudication layer over the validated candidates. \textsf{BioAgent}, \textsf{ChemAgent}, and \textsf{RiskAgent} receive the same compact evidence summary and return claim $\rightarrow$ evidence $\rightarrow$ impact critiques. \textsf{DecisionAgent} then synthesizes these critiques into a final modality decision and an adjusted ranking, while remaining restricted to evidence already present in the context. Representative prompt excerpts and training examples appear in Appendix~\ref{app:prompts}.

\subsection{Scoring, Risk Flags, and Audit Log}
Conceptually, the candidate-ranking logic can be viewed as following a mode-specific heuristic of the form
\[
S_0(r) = s_{\textsf{mode}}(r) - p_{\textsf{TM}}(r) - p_{\textsf{PTM}}(r) - p_{\textsf{motif}}(r),
\]
where $s_{\textsf{mode}}$ captures the base preference, while the penalty terms summarize constraint-aware adjustments from TM overlap, PTM-mask overlap, and motif context.  In particular, epitope scoring favors polar, non-TM, PTM-light windows, whereas pocket scoring places more weight on hydrophobicity and applies weaker PTM penalties, with small motif-specific adjustments in some cases. In our inference pipeline, the \LLM\ proposes and ranks candidates using this evidence-rich context.

We also compute a separate risk vector $g(r)$ producing typed flags such as TM-overlap, PTM-overlap, glyco-mask-overlap, PTM-dense, disulfide-constrained, hydrophobic-core, and motif-overlap. These flags are logged as interpretable annotations alongside the heuristic score. The detailed rules that determine the flags are noted in Appendix~\ref{app:hydropathy}--\ref{app:feature_ref}.

\section{Experimental Results}

Evaluating AI agents for drug-target site discovery is inherently challenging because there is no widely accepted benchmark for this task. In addition, for a novel target-site discovery, there is no definitive ``ground truth". We therefore evaluate \sysname{} using a comprehensive, modality-aware validation data rather than a single fixed benchmark. 

Our evaluation has three goals: (i) to test whether the predicted regions overlap known \textit{small-molecule pockets}, (ii) to assess whether predicted epitope regions agree with annotated \textit{antibody epitope}s, and (iii) to examine whether predicted sites remain \textit{structurally plausible} even when \sysname{} does not directly use protein structure as input.

\begin{table}[htbp]
\centering
\caption{Overview of the curated datasets used for final reporting}
\label{tab:curated_dataset_overview}
\setlength{\tabcolsep}{5pt}
\renewcommand{\arraystretch}{1.0}
\begin{tabular}{lcc}
\hline
\textbf{Drug Modality} & \textbf{Count} \\
\hline
Small molecule (S) & 55 \\
Antibody (A) & 26  \\
Antibody / Small molecule (AS) & 8 \\
\hline
Total & 89 \\
\hline
\end{tabular}
\end{table}

To support this evaluation, we curated a dataset spanning pocket, epitope, and mixed-modality targets. Table~\ref{tab:curated_dataset_overview} summarizes the final reporting groups. Here, \texttt{Group S} denotes pocket-mode small-molecule validation cases, \texttt{Group A} denotes antibody epitope validation cases, and \texttt{Group AS} denotes mixed-modality targets that are known to support both pocket and epitope interpretations. This organization reflects the fact that therapeutic site selection is modality-dependent and that different sources provide different levels of supervision.


\subsection{Assessment on small-molecule targets (Pockets)}

\begin{figure*}[!t]
  \centering

  \begin{subfigure}[t]{0.48\textwidth}
    \centering
    \includegraphics[width=\textwidth]{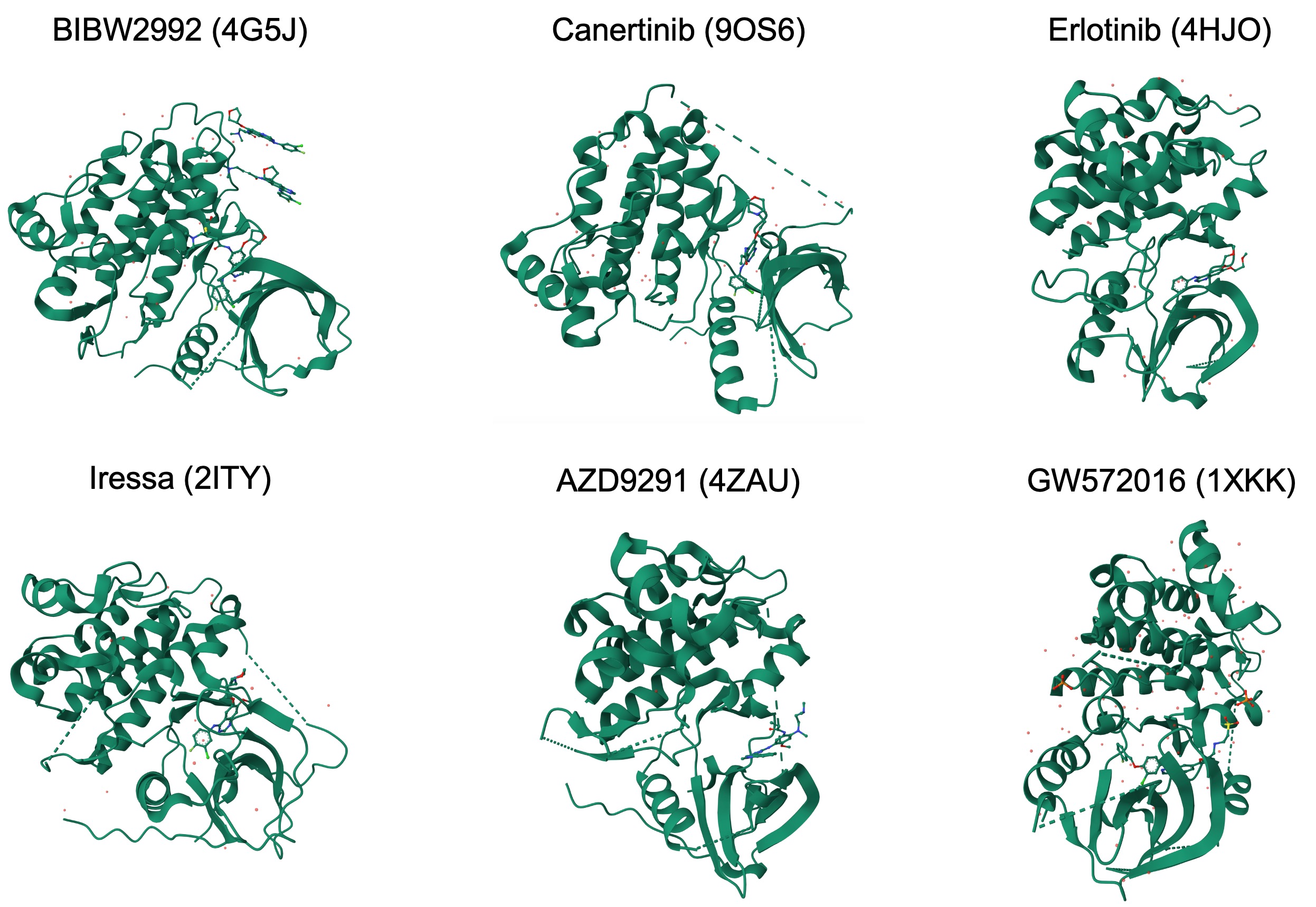}
    \caption{Six EGFR--drug cocrystal structures with similar orientations.}
    \label{fig:egfr_pockets}
  \end{subfigure}
  \hfill
  \begin{subfigure}[t]{0.48\textwidth}
    \centering
    \includegraphics[width=\textwidth]{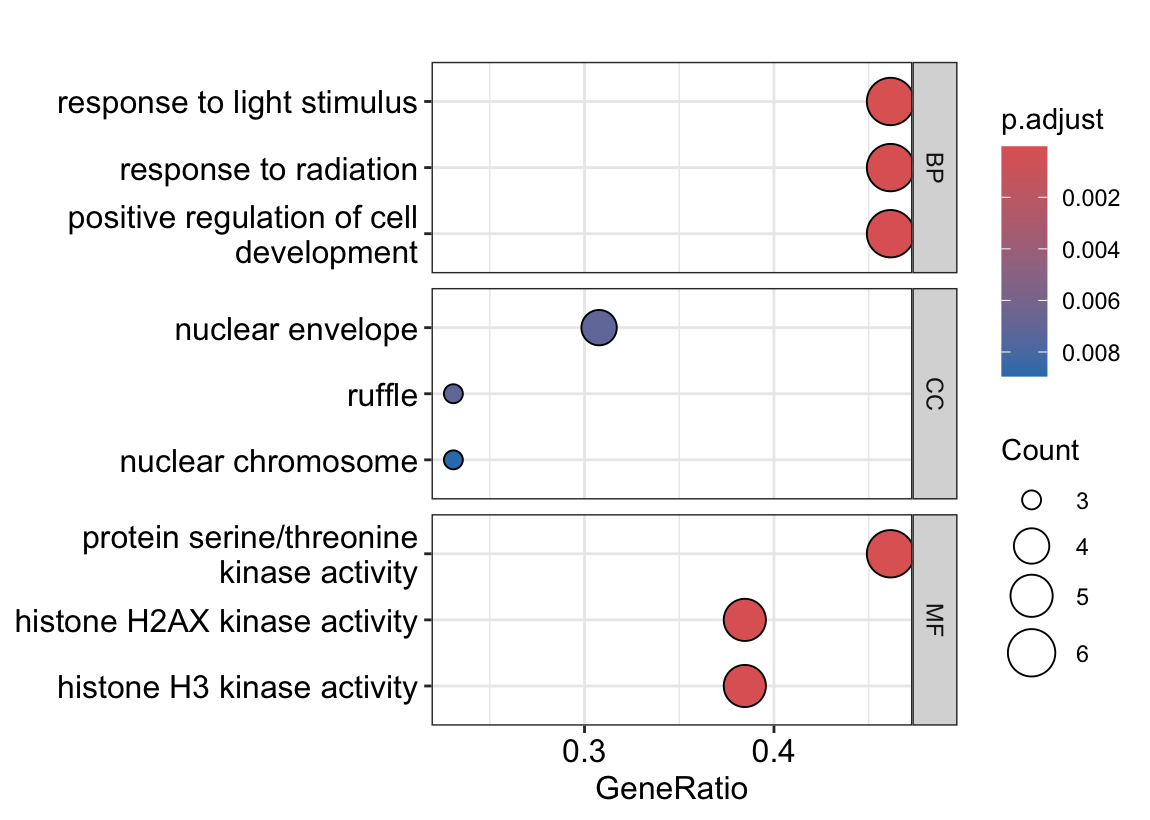}
    \caption{GO plot of targets of interest from Site4Drug.}
    \label{fig:go_plot_targets}
  \end{subfigure}

  \vspace{0.8em}

  \begin{subfigure}[t]{0.58\textwidth}
    \centering
    \includegraphics[width=\textwidth]{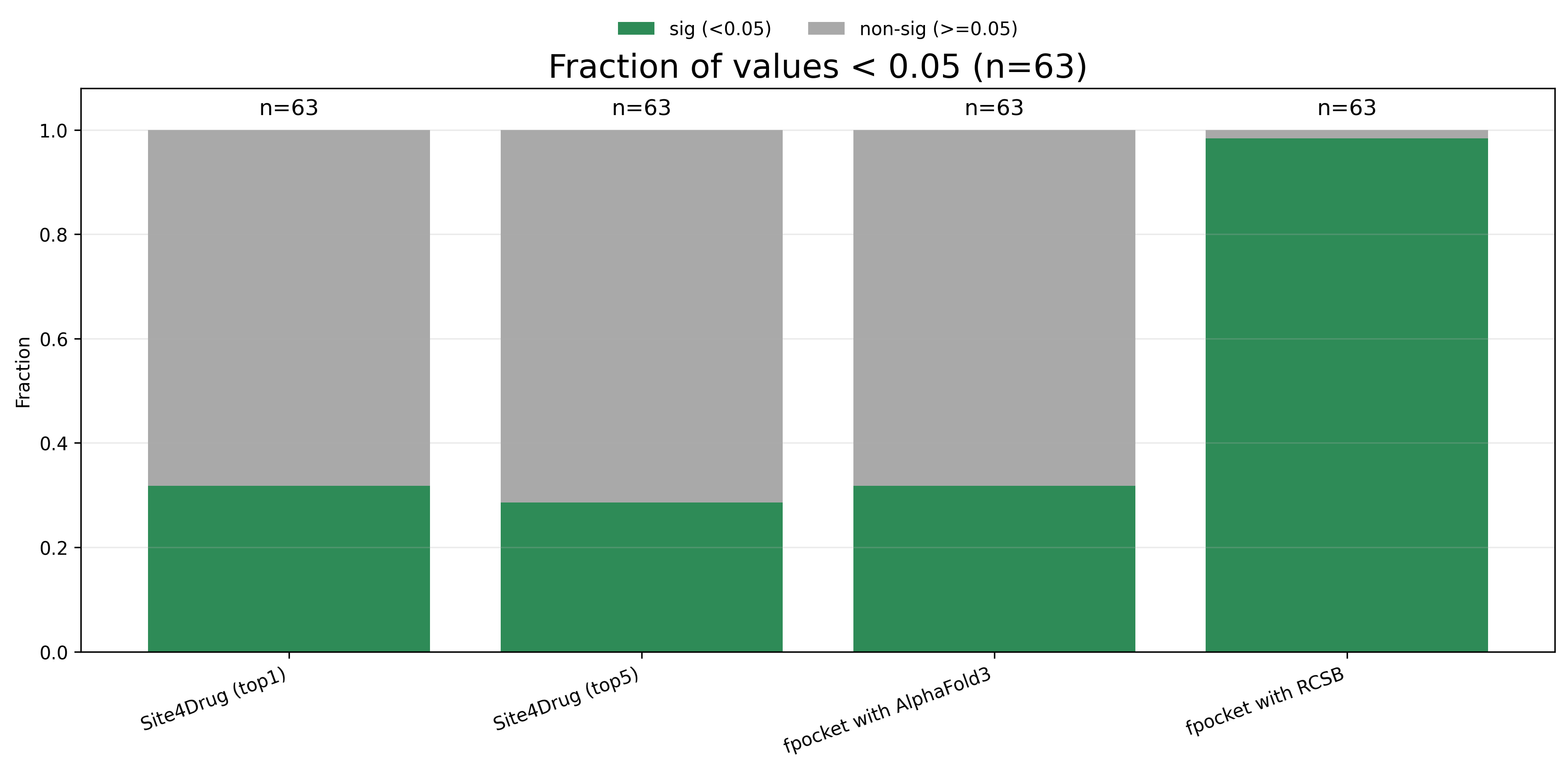}
    \caption{Proportion of targets with pocket site predictions at $p$-value $< 0.05$ for each baseline method. }
    \label{fig:baseline_sig_ratio}
  \end{subfigure}
  \begin{subfigure}[t]{0.38\textwidth}
    \centering
    \includegraphics[width=\textwidth]{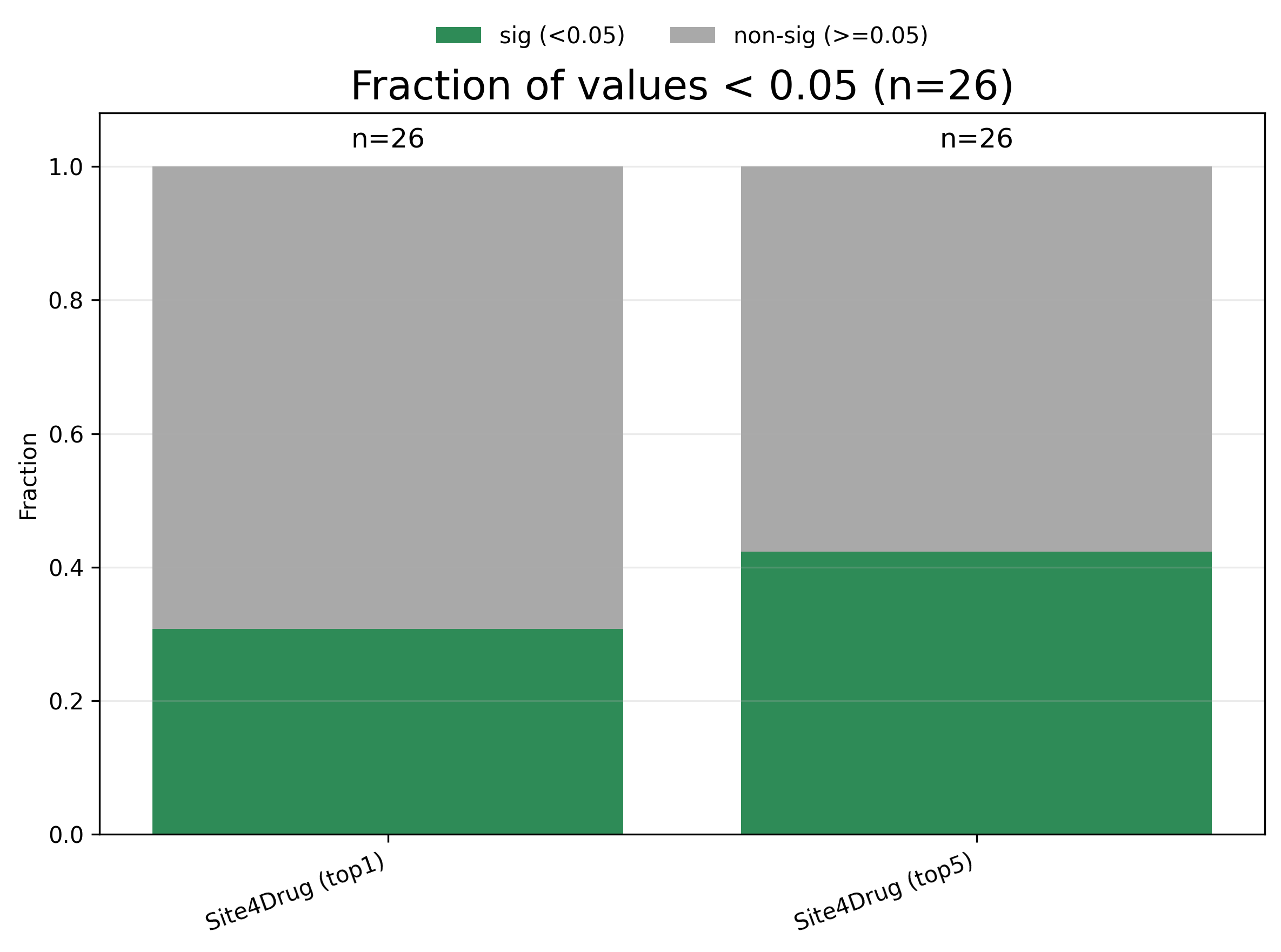}
    \caption{Proportion of targets with epitope site predictions at $p$-value $< 0.05$. }
    \label{fig:baseline_sig_ratio_epitope}
  \end{subfigure}

  \caption{Overview of structural examples, functional enrichment results, and baseline significance ratios. (a) Six EGFR--drug cocrystal structures viewed from similar orientations along with the corresponding drugs and RCSB entries. (b) GO plot of targets for which the top-1 Site4Drug LLM prediction achieved $p$-value $< 0.05$. (c) The proportion of targets with significant site predictions ($p$-value $< 0.05$) on the pocket benchmark ($n=63$): Site4Drug top-1 (20/63), Site4Drug top-5 (18/63), fpocket with AlphaFold3 input (20/63), and fpocket with RCSB input (62/63). \sysname{} produced comparable results to fpocket with AlphaFold3 structure, showing \sysname{} implies structural information under general conditions. Since the RCSB structure contains the bound ligand and thus adopts a bent conformation, fpocket with ligand-bound RCSB structures yielded results very close to the ground truth (98.41\%) as expected. (d) Applying \sysname{} to the ABCD database with epitope mode yielded a similar proportion of targets with significant site predictions ($p$-value $< 0.05$) as in pocket mode.}
  \label{fig:egfr_baseline_go}
\end{figure*}

We first evaluated \sysname{} on small-molecule targets for which experimentally determined target--drug co-crystal structures (from methods such as cryo-EM or X-ray crystallography) are available. In this setting, we use deposited structures from public resources (i.e., RCSB) to define approximate ground-truth pocket residues. Specifically, for each target--drug complex, the pocket is defined as the set of residues located within 4~\AA\ of the bound ligand. These residue indices are then remapped from structure numbering to the input FASTA sequence using Needleman--Wunsch alignment. This yields a sequence-level reference site that can be directly compared against the sequence-defined outputs of \sysname{}.
Fig.~\ref{fig:egfr_pockets} shows examples of co-crystals of EGFR and six different small molecule ligands. This visually shows that our validation data is composed of pocket locations that do not differ significantly among small molecule drugs, making it ideal for evaluation purposes.

Following the procedure above, we initially collected 91 literature-based target--small-molecule pairs with corresponding RCSB entries and retained (\texttt{Group S} + \texttt{Group AS}) 63  after excluding cases in which a representative FASTA sequence could not be identified reliably. For example, this is the case of multi-chain ambiguity or difficult residue mapping. Examples of representative validation data are shown in Table~\ref{tab:rcsb_cocrystal_examples}. 

\begin{table*}[!t]
\centering
\caption{Representative entries from the RCSB target--drug co-crystal dataset.
\texttt{Site (FASTA)} denotes binding-site residues located within 4~\AA\ of the co-crystallized drug, remapped onto the input FASTA coordinate.}
\label{tab:rcsb_cocrystal_examples}
\footnotesize
\begin{tabular}{@{}llllp{7.1cm}p{2.8cm}@{}}
\toprule
Target & Drug & Mode & PDB ref. & Site (FASTA) & Description \\
\midrule
CYP17A1 & Abiraterone & pocket & \texttt{3RUK} &
95, 96, 183, 184, 187, 191, 279, 280, 284, 288, 348, 424, 464 &
Decreases PSA. \\

AKT1 & Borussertib & pocket & \texttt{6HHF} &
17, 54, 79, 80, 84, 85, 210, 211, 264, 268, 271, 272, 273, 274, 290, 292, 296, 297 &
Covalent-allosteric. \\

KIF11 (Eg5) & Ispinesib & pocket & \texttt{4A5Y} &
116, 117, 118, 119, 127, 130, 133, 136, 137, 160, 211, 214, 215, 217, 218, 221 &
Inhibits KSP. \\
\bottomrule
\end{tabular}
\end{table*}

\begin{table*}[t]
\centering
\caption{Representative target--antibody pairs from the ABCD database with available epitope site annotations.}
\label{tab:abcd_epitope_examples}
\scriptsize
\resizebox{\textwidth}{!}{%
\begin{tabular}{@{}lll p{6.8cm} l@{}}
\toprule
Target & Antibody & Mode & Epitope site & Reference \\
\midrule
\texttt{ADAM17} & \texttt{MEDI3622} & epitope & \texttt{PKAYYSPVGKKNIYLN; 366-381} & \texttt{ABCD\_AR537} \\
\texttt{malE / MBP} & \texttt{alpha-MBP-scFv} & epitope & \texttt{ELAKKFEKDTGIKV; 51-61} \newline \textit{(not including ELA at the front)} & \texttt{ABCD\_AW485} \\
\texttt{F} & \texttt{MPE8} & epitope & \texttt{SKGYLSALRTGWYTS + QLPLYGVIDT; 41-55, 302--311} \newline \textit{(neutralizes HRSV and HMPV isolates of several subgroups)} & \texttt{ABCD\_AR272} \\
\bottomrule
\end{tabular}%
}
\end{table*}

 Then, we ran \sysname{} on this data and first verified that the curated proteins were classified as pocket. A few targets, including EGFR and HER2, were classified as epitope modalities by \sysname{}. However, these were acceptable cases that are known to have both corresponding antibody and small molecule drugs.\footnote{EGFR is well known for several small-molecule ligands, including afatinib, canertinib, erlotinib, gefitinib, lapatinib, and osimertinib, making it a pocket-mode case. At the same time, it is also targeted by antibodies such as cetuximab, panitumumab, nimotuzumab, and 806, which makes it a natural mixed-modality test case. Meanwhile, HER2 is also associated with both small-molecule drugs (e.g., lapatinib, neratinib, sapitinib, and CI-1033) and antibody drugs (e.g., margetuximab, trastuzumab, disitamab, and pertuzumab).}  Proceeding with actual pocket binding site evaluation, we compared the top-1 and top-5 sites predicted by \sysname{} against the reference pocket residues by performing hypergeometric distribution-based test. The results for cases with p-values below 0.05 are reported in Fig.~\ref{fig:baseline_sig_ratio}. We additionally compared \sysname{} against fpocket as a structure-based baseline. Structure-trained models such as RAPID-Net were excluded from the main comparison because they were trained on scPDB, which is derived from RCSB and would therefore introduce a substantial risk of data leakage in this benchmark.

Under this evaluation, \sysname{} achieved pocket-site localization performance comparable to fpocket under general conditions even without direct structure information (Fig.~\ref{fig:baseline_sig_ratio}). We also discovered that the statistically significant top-1 results were mainly kinases, which is biologically plausible because kinases often contain characteristic and recurrent small-molecule-accessible pockets (Fig.~\ref{fig:go_plot_targets}). For instance, pralsetinib acts as a multi-kinase inhibitor that targets 11 kinases: DDR1, FGFR1, FGFR2, FLT3, JAK1, JAK2, KDR, NTRK1, NTRK3, PDGFRB, and RET \cite{gonzalez2025combinatorial}. As expected, fpocket applied directly to ligand-bound RCSB structures achieved near-ceiling performance because the bound ligand is already present and strongly constrains the pocket geometry (orientation leaks the answer). To complement the binary $p < 0.05$ criterion, we provide the full prediction outputs, reference pocket residues, and raw top-1/top-5 hypergeometric p-values for all evaluated cases in the GitHub repository, enabling readers to examine non-significant predictions in more detail. The main value of \sysname{} is that it can still propose meaningful sites when such structural context is absent, incomplete, ambiguous, or difficult to use directly.

\subsection{Ablation: Sequence-only baseline}

To test whether \sysname{} benefits from its explicit evidence pipeline beyond sequence-only domain-aligned prompting, we constructed a sequence-only ablation on the same included pocket benchmark used in the main small-molecule evaluation (\texttt{Group S + Group AS}, $n=63$). We chose this setting because it is the largest and cleanest benchmark in this study and also supports direct comparison with the structure-based baseline fpocket. In this ablation, the model received only a generic target identifier and the raw amino-acid sequence. Otherwise, the prompt structure, inference setting, and output format were kept nearly identical to the main pipeline, with the explicit TM/topology, PTM, motif, and cysteine/disulfide evidence fields removed. Exact prompt details are provided in Appendix~\ref{app:seq_only_prompt}.
We also assessed the effect of self-consistency by comparing a single proposal (\texttt{$k=1$}) against three-attempt voting (\texttt{$k=3$}). The sequence-only baseline yielded significant overlap ($p<0.05$) for 3/63 cases at both top-1 and top-5 under \texttt{$k=1$}, improving to 7/63 and 6/63 under \texttt{$k=3$}, respectively. These values remain well below the full \sysname{} pipeline (Fig.~\ref{fig:baseline_sig_ratio}), suggesting that \sysname{} benefits materially from its explicit evidence pipeline rather than merely exploiting generic sequence-level patterns.

\subsection{Assessment on antibody targets (Epitope)}

We next evaluated \sysname{} in epitope mode using antibody--target pairs curated from the AntiBodies Chemically Defined (ABCD) database. In contrast to the pocket benchmark above, residue-level epitope annotations were much sparser in this setting: many ABCD entries identify the target and antibody but do not provide a precise sequence-resolved epitope. We therefore retain only the subset with sufficiently clear annotated epitope positions for direct overlap-based evaluation. In total, we initially curated 52 target--antibody pairs and retained 26 for final epitope benchmarking \texttt{Group A}. Representative examples are shown in Table~\ref{tab:abcd_epitope_examples}. 

For each retained sequence, we evaluated whether the top-1 or top-5 predictions from \sysname{} significantly overlapped the annotated epitope using the same hypergeometric framework as in pocket mode. When applied in epitope mode, \sysname{} produced significant overlap ($p < 0.05$) for 8 cases under the top-1 criterion and for 11 cases under the top-5 criterion (Fig.~\ref{fig:baseline_sig_ratio_epitope}). Although this benchmark is much smaller and noisier than the small-molecule benchmark, the observed success rate indicates that \sysname{} can recover meaningful epitope-like regions from sequence-derived evidence.

\subsection{Validation of Structural Reliability}

Finally, we asked whether the sites predicted by \sysname{} tend to fall in structurally credible regions even though the model does not directly ingest protein structure. For this analysis, we used AlphaFold3 \cite{abramson2024alphafold3} to predict structures corresponding to the 63 retained pocket-mode validation target pairs and examined the \pLDDT{} values of the residues contained in the predicted sites.

Regions with lower \pLDDT{} are generally interpreted as less structurally reliable or more conformationally flexible, whereas higher-\pLDDT{} regions tend to correspond to more confidently modeled local structure. Although the drug-binding site does not have to lie exclusively in the highest-\pLDDT{} region of a protein, structural confidence remains a useful secondary plausibility check, especially for novel sites that lack direct experimental annotation.

\begin{figure*}[!t]
  \centering
  \includegraphics[width=0.4\textwidth]{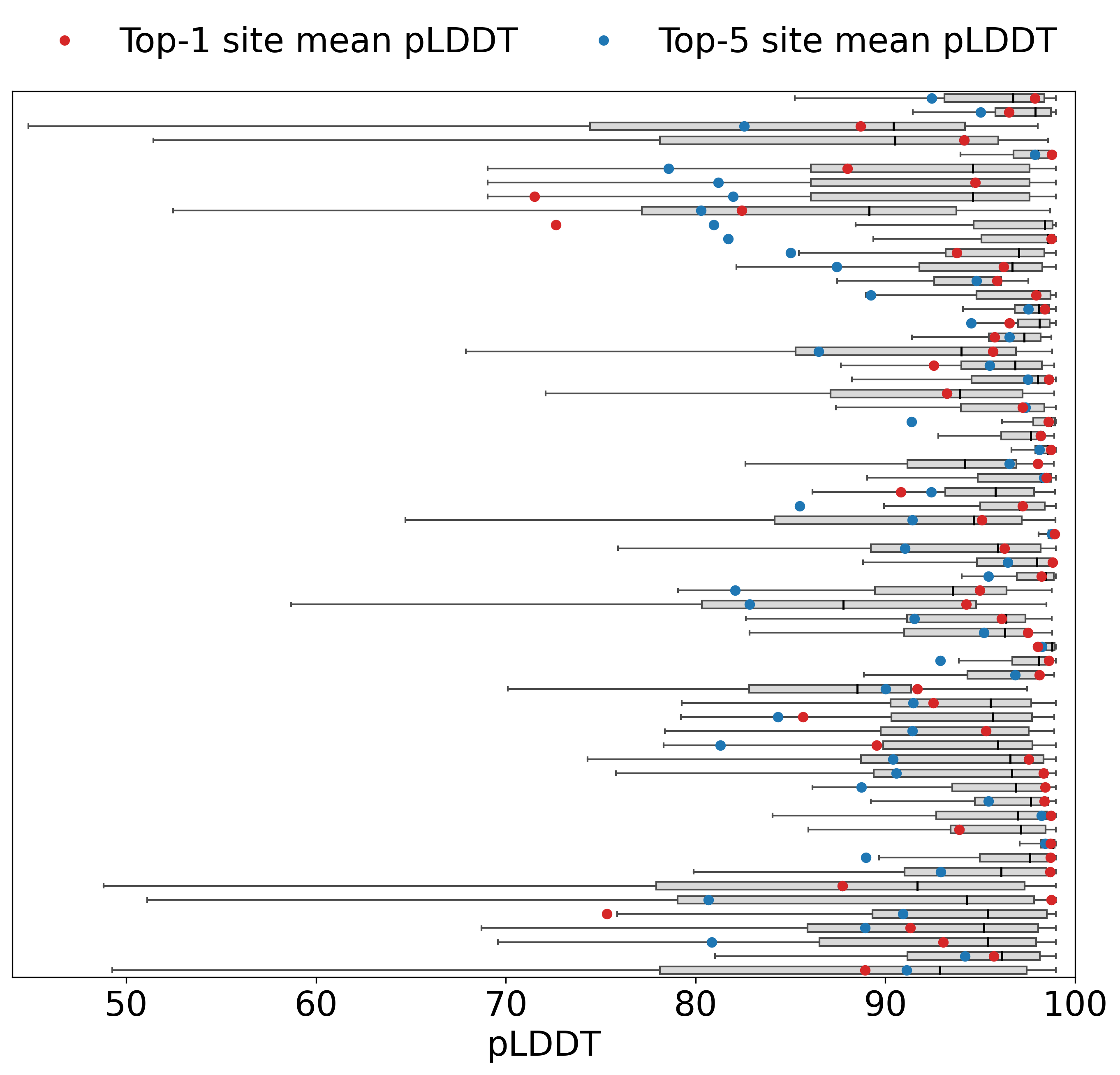}
  \caption{
  Horizontal box plot of per-residue \pLDDT{} values for the corresponding AlphaFold3 structures of 63 records included in pocket-mode validation (\texttt{Group S} and \texttt{Group AS} groups). 
  Each box plot summarizes the whole-structure \pLDDT{} distribution for one record. 
  The red marker denotes the mean \pLDDT{} of the best site, while the blue marker denotes the mean \pLDDT{} averaged over top-5 annotated sites in that record. 
  Together, these markers show how local site confidence compares with the overall structural confidence of each predicted site.
  }
  \label{fig:combined_boxplot}
\end{figure*}

To examine whether the sites predicted by \sysname{}, despite not using structural information, still correspond to structurally reliable regions that are favorable for drug binding, we compared the \pLDDT{} values of \sysname{}’s top-1 and top-5 predictions on the AlphaFold3 structures corresponding to the 63 cases used in the pocket validation. In most cases, the top-1 site had a higher mean \pLDDT{} than the broader top-5 set, with only 9 exceptions (Fig.~\ref{fig:combined_boxplot}). This suggests that the highest-ranked prediction from \sysname{} is often not only statistically meaningful with respect to known sites, but also structurally more plausible than lower-ranked alternatives. More broadly, this behavior indicates that the sequence-based evidence aggregation used by \sysname{} may implicitly recover signals correlated with structural reliability, even without direct structure input.

\section{Module 2: Modality-specific Design Handoff}
\textbf{Module 1 addresses the challenge of selecting \emph{where} to intervene on a protein (i.e., choosing a targetable site}). Thus, ranked outputs of Module~1 can now be treated by Module~2 as an interface to downstream design tools. In epitope mode, top-ranked spans together with topology/PTM risk annotations can be passed to peptide or antibody-binder design tools (e.g., BoltzGen). In pocket mode, top-ranked pocket regions lead to small-molecule design or scoring tools (e.g., DrugCLIP, BindCLIP). 

Thus, we also expand and report the capability of not only providing the target site, but also end-to-end design. While the following results illustrate the potential of Module~2 for end-to-end design workflows, they should be interpreted as proof-of-concept demonstrations on a single target, EGFR, rather than as broad validation across diverse targets.

\subsection{Application of Module 2 for pocket mode.}



Because DrugCLIP takes as input a three-dimensional structure that can be defined as a pocket within the full protein structure, we extracted only the atoms belonging to the predicted site from the AlphaFold3 structure and used them to construct the pocket. DrugCLIP was then applied to the top-ranked site predicted by Site4Drug for EGFR (RCSB code: 1XKK; p-value $= 8.47 \times 10^{-5}$). Known EGFR inhibitors including afatinib, canertinib, erlotinib, gefitinib, lapatinib, and osimertinib were absent from the DrugCLIP ligand database and therefore could not appear in the retrieved list. However, visualization of the top-6 hits together with these reference compounds showed that the retrieved molecules were structurally similar to the EGFR inhibitors and shared structural motifs likely to bind this pocket, such as benzene and cyclohexane moieties (Tab. \ref{tab:top6_smiles}, Fig. \ref{fig:drugclip_egfr}). This observation was further supported by a hypergeometric test evaluating the overlap between the structures of the top-6 hits and the ground truth (lapatinib; 1XKK) site, which yielded p-value lower than $10^{-11}$ (almost identical). For this analysis, we used Boltz2 \cite{passaro2025boltz2}, because AlphaFold3 does not currently support general ligand-containing predictions.

\subsection{Application of Module 2 for epitope mode.}
We ran BoltzGen using the top-1 to top-5 EGFR epitopes predicted by \sysname{} to generate peptide binders. Because BoltzGen tends to encounter memory issues with protein sequences longer than approximately 1,000 amino acids, we truncated the 1,210-aa EGFR precursor to retain only the first 1,000 aa from the N terminus. We then provided the epitope information predicted by \sysname{} and ran BoltzGen with the number of designs set to 50 and the peptide length constrained to 12- to 18-mers. Among the generated candidates, the binder with the best BoltzGen score was selected as the representative peptide binder for each epitope. 

The peptide and EGFR sequences were then provided to AlphaFold3 to generate five multimer structures, from which we calculated the median Local Interaction Score (LIS) \cite{kim2024enhanced} and the median of the minimum distance between the peptide and the target (Fig. \ref{fig:boltzgen}).

It is straightforward to understand why rank-5 received the lowest priority in \sysname{}, as unlike the higher-priority sites, it is located intracellularly. Predicted aligned error (PAE) originally indicates higher confidence in the relative positioning of two residues when its value is lower. Using the commonly applied cutoff PAE value of 12 for LIS calculation, only rank-1 yielded a definable LIS, whereas all PAE values for the remaining extracellular sites exceeded 12. The high ranking of rank-2 is also associated with the fact that it corresponds to Domain III of EGFR, which is known to be targeted by antibody therapeutics such as petosemtamab and 7D12. For rank-3 and rank-4, a difference was observed in the minimum distance metric. 

These results show that the final binder candidates generated from \sysname{}'s site ranking exhibit consistency with prior literature and structural information. 

\section{Discussion and Limitations}

\textbf{SFT-centric supervision.}
We initially considered supervised fine-tuning (SFT) on structured demonstrations as a post-training strategy. An example is shown in Appendix~\ref{app:format_shift}. In preliminary experiments with the Qwen backbone (Qwen3-235B Instruct), the SFT checkpoint improved output formatting but also showed shortcut behavior, including repeated selection of similar N-terminal spans. We therefore report quantitative results using the base model. This suggests that post-training for target-site discovery may require biologically grounded reward or preference signals, which we view as an important direction for future work.

\textbf{Structure is sometimes gating.}
Especially, pocket mode depends on structural context, and some models take structural information as input. 
However, this not only requires a large number of tokens (e.g., more than 5,000 amino acids do not provide structure in AlphaFold3) but is also not suitable for scalable screening purposes. Therefore, it is important to use an appropriate level of structural information. 

\textbf{No consideration of quaternary structure.}
At present, the model takes a single protein sequence as input, but in reality multiple proteins can also assemble into a quaternary structure. In fact, many drugs commonly target channel proteins composed of multiple domains. It will therefore be necessary to extend the approach to handle such cases as well. Unlike fpocket and RAPID-Net, which take single amino acids as input, the LLM-based Site4Drug, capable of leveraging diverse metadata including pocket / epitope mode designation, appears well positioned to address this cochain issue with ease.

\textbf{More available information.}
Predictions may become more accurate as broader databases and annotations are integrated.
For example, curated information such as exact disulfide-bond annotations from \Uniprot\ is not yet used throughout the present pipeline.

\textbf{Sensitivity to topology.} It should be noted that, rather than inputting the full amino acid sequence, partial sequence input may also be used to obtain a more context-specific epitope or to avoid token limit constraints. However, caution is warranted in such cases, as the topology may be inferred in the reverse orientation.

\textbf{Concentration dependency.}
Target engagement is concentration-dependent, as some compounds are inactive at lower concentrations but exhibit measurable interactions with specific targets at higher concentrations. We need to test Site4Drug on data that includes concentration context.

\textbf{Scale and breadth of validation and baseline comparison.} 
A key limitation of this study is the limited scale of validation and fair baseline comparison. Although Site4Drug was evaluated using modality-aware validation datasets, the current datasets remain modest in size, particularly for antibody-relevant target sites where reliable site-level annotations are difficult to curate. Moreover, direct comparison with existing structure-trained machine learning models is challenging because of potential data leakage from overlapping training databases. Future work should address these limitations by expanding validation to larger and more diverse targets under careful leakage control, and by incorporating fair apo-structure-based or sequence-based ML baselines that are independent of the evaluation set.

\section{Conclusion}
\sysname\ aims to improve the reliability and transparency of target-site selection by producing constraint-aware region proposals and audit-ready decision logs. Risks include over-reliance on incomplete evidence sources or misinterpretation of immune epitope data. We mitigate these risks through explicit uncertainty flags and traceable audit trails. By identifying targetable protein regions together with the evidence supporting each decision, \sysname{} provides an interpretable upstream interface for downstream binder or small-molecule design workflows. For example, we applied \sysname{} for diverse drug modalities: small molecules, antibodies, and peptide drugs. We also tested that our site detection can be easily combined with third-party tools, such as Diffuse Bio Sandbox, a platform for nanobody and scFv modeling (Fig.~\ref{fig:egfr_diffusebio_appendix}).

We have confirmed that the epitope location predicted by \sysname{} matched exactly with the Hydrogen-Deuterium Exchange Mass Spectrometry result for an actual target. Interestingly, antibody physicochemical properties affected target binding, and our pipeline suggested a possible mechanism (Appendix~\ref{app:failuremode}). Additional experiments (e.g., alanine scanning) on multiple targets and high-throughput screening tests are planned for future studies. Additionally, future research may focus on a lab-in-the-loop framework for \sysname{}, integrating real MS data and iterative experimental feedback.
Furthermore, \sysname{} will be incorporated into automated target-discovery pipelines \cite{zhang2026virtualbiotech}, thereby enabling end-to-end applications.

\section*{Impact Statement}
This work aims to advance machine learning methods for drug-target site discovery and to support downstream small-molecule, peptide, and antibody-binder design. The potential positive impact is to accelerate early-stage therapeutic research by helping prioritize candidate binding sites, particularly for targets with limited experimental information. However, because such predictions may be used together with generative design tools, responsible use requires caution. Site4Drug outputs should be treated as hypotheses for experimental follow-up, not as validated therapeutic candidates. Any downstream molecules designed from these predictions should undergo independent validation for binding, specificity, off-target effects, safety, and regulatory suitability before translational or clinical use.

\section*{Acknowledgements}
We implemented the LLM-agent pipeline using the Tinker API at Thinking Machines Lab. This project was developed as an extension of coursework in CDSS 94 at UC Berkeley, and we thank Kevin Miao and Karina Nguyen for course support and guidance. We also thank Diffuse Bio for providing sandbox credits. The corresponding author, Jeongbin Park, gained expertise in in silico drug design while working at Portrai, Inc.

\section*{Code Availability}
The demo for \sysname{} is available at \url{https://github.com/winterrykim/Site4Drug\_Demo}.


\bibliography{example_paper}

@article{stark2025boltzgen,
  title={Boltzgen: Toward universal binder design},
  author={Stark, Hannes and Faltings, Felix and Choi, MinGyu and Xie, Yuxin and Hur, Eunsu and O’Donnell, Timothy and Bushuiev, Anton and U{\c{c}}ar, Talip and Passaro, Saro and Mao, Weian and others},
  journal={bioRxiv},
  pages={2025--11},
  year={2025},
  publisher={Cold Spring Harbor Laboratory}
}

@article{gao2023drugclip,
  title={Drugclip: Contrastive protein-molecule representation learning for virtual screening},
  author={Gao, Bowen and Qiang, Bo and Tan, Haichuan and Jia, Yinjun and Ren, Minsi and Lu, Minsi and Liu, Jingjing and Ma, Wei-Ying and Lan, Yanyan},
  journal={Advances in Neural Information Processing Systems},
  volume={36},
  pages={44595--44614},
  year={2023}
}

@article{qiao2026bindclip,
  title={BindCLIP: A Unified Contrastive-Generative Representation Learning Framework for Virtual Screening},
  author={Qiao, Anjie and Wang, Zhen and Li, Yaliang and Rao, Jiahua and Yang, Yuedong},
  journal={arXiv preprint arXiv:2602.15236},
  year={2026}
}

@article{vita2025immune,
  title={The immune epitope database (IEDB): 2024 update},
  author={Vita, Randi and Blazeska, Nina and Marrama, Daniel and IEDB Curation Team Members Shackelford Deborah Zalman Leora Foos Gabriele Zarebski Laura Chan Kenneth Reardon Brian Fitzpatrick Sidne Busse Matthew Coleman Sara Sedwick Caitlin Edwards Lindy MacFarlane Catriona Ennis Marcus and Duesing, Sebastian and Bennett, Jason and Greenbaum, Jason and De Almeida Mendes, Marcus and Mahita, Jarjapu and Wheeler, Daniel K and others},
  journal={Nucleic Acids Research},
  volume={53},
  number={D1},
  pages={D436--D443},
  year={2025},
  publisher={Oxford University Press}
}

@article{le2009fpocket,
  title={Fpocket: an open source platform for ligand pocket detection},
  author={Le Guilloux, Vincent and Schmidtke, Peter and Tuffery, Pierre},
  journal={BMC bioinformatics},
  volume={10},
  number={1},
  pages={168},
  year={2009},
  publisher={Springer}
}

@article{balytskyi2025rapid,
  title={RAPID-Net: Accurate Pocket Identification for Binding-Site-Agnostic Docking},
  author={Balytskyi, Yaroslav and Hubenko, Inna and Balytska, Alina and Kelly, Christopher V},
  journal={Journal of Chemical Information and Modeling},
  volume={65},
  number={22},
  pages={12221--12239},
  year={2025},
  publisher={ACS Publications}
}

@article{kyte1982simple,
  title={A simple method for displaying the hydropathic character of a protein},
  author={Kyte, Jack and Doolittle, Russell F},
  journal={Journal of molecular biology},
  volume={157},
  number={1},
  pages={105--132},
  year={1982},
  publisher={Elsevier}
}

@article{wang2020musitedeep,
  title={MusiteDeep: a deep-learning based webserver for protein post-translational modification site prediction and visualization},
  author={Wang, Duolin and Liu, Dongpeng and Yuchi, Jiakang and He, Fei and Jiang, Yuexu and Cai, Siteng and Li, Jingyi and Xu, Dong},
  journal={Nucleic Acids Research},
  volume={48},
  number={W1},
  pages={W140--W146},
  year={2020},
  publisher={Oxford University Press}
}

@article{de2006scanprosite,
  title={ScanProsite: detection of PROSITE signature matches and ProRule-associated functional and structural residues in proteins},
  author={De Castro, Edouard and Sigrist, Christian JA and Gattiker, Alexandre and Bulliard, Virginie and Langendijk-Genevaux, Petra S and Gasteiger, Elisabeth and Bairoch, Amos and Hulo, Nicolas},
  journal={Nucleic acids research},
  volume={34},
  number={suppl\_2},
  pages={W362--W365},
  year={2006},
  publisher={Oxford University Press}
}

@article{cornelio2023ai_descartes,
  title={Combining data and theory for derivable scientific discovery with AI-Descartes},
  author={Cornelio, Cristina and Dash, Sanjeeb and Austel, Vernon and Josephson, Tyler R and Goncalves, Joao and Clarkson, Kenneth L and Megiddo, Nimrod and El Khadir, Bachir and Horesh, Lior},
  journal={Nature Communications},
  volume={14},
  number={1},
  pages={1777},
  year={2023},
  publisher={Nature Publishing Group UK London}
}

@article{jansen2026literature,
  title={Generating Literature-Driven Scientific Theories at Scale},
  author={Jansen, Peter and Clark, Peter and Downey, Doug and Weld, Daniel S},
  journal={arXiv preprint arXiv:2601.16282},
  year={2026}
}

@article{agarwal2026autodiscovery,
  title={Autodiscovery: Open-ended scientific discovery via bayesian surprise},
  author={Agarwal, Dhruv and Majumder, Bodhisattwa Prasad and Adamson, Reece and Chakravorty, Megha and Gavireddy, Satvika Reddy and Parashar, Aditya and Surana, Harshit and Dalvi Mishra, Bhavana and McCallum, Andrew and Sabharwal, Ashish and others},
  journal={Advances in Neural Information Processing Systems},
  volume={38},
  pages={25181--25219},
  year={2026}
}

@article{swanson2024virtuallab,
  title={The virtual lab: Ai agents design new sars-cov-2 nanobodies with experimental validation. bioRxiv, 2024},
  author={Swanson, K and Wu, W and Bulaong, NL and Pak, JE and Zou, J},
  journal={bioRxiv},
  volume={11},
  year={2024}
}

@article{gottweis2026cscientist,
  title={Accelerating scientific discovery with Co-Scientist},
  author={Gottweis, Juraj and Weng, Wei-Hung and Daryin, Alexander and Tu, Tao and Sirkovic, Petar and Myaskovsky, Artiom and Glowaty, Grzegorz and Weissenberger, Felix and Orlandi, Alessio and Popovici, Dan and others},
  journal={Nature},
  pages={1--3},
  year={2026},
  publisher={Nature Publishing Group UK London}
}

@article{zhang2026virtualbiotech,
  title={The Virtual Biotech: A Multi-Agent AI Framework for Therapeutic Discovery and Development},
  author={Zhang, Harrison G and Eckmann, Peter and Miao, Jiacheng and Mahon, Andrew B and Zou, James},
  journal={bioRxiv},
  pages={2026--02},
  year={2026},
  publisher={Cold Spring Harbor Laboratory}
}

@article{passaro2025boltz2,
  title={Boltz-2: Towards accurate and efficient binding affinity prediction},
  author={Passaro, Saro and Corso, Gabriele and Wohlwend, Jeremy and Reveiz, Mateo and Thaler, Stephan and Somnath, Vignesh Ram and Getz, Noah and Portnoi, Tally and Roy, Julien and Stark, Hannes and others},
  journal={BioRxiv},
  year={2025}
}

@article{lu2026towards,
  title={Towards end-to-end automation of AI research},
  author={Lu, Chris and Lu, Cong and Lange, Robert Tjarko and Yamada, Yutaro and Hu, Shengran and Foerster, Jakob and Ha, David and Clune, Jeff},
  journal={Nature},
  volume={651},
  number={8107},
  pages={914--919},
  year={2026},
  publisher={Nature Publishing Group UK London}
}

@article{abramson2024alphafold3,
  title={Accurate structure prediction of biomolecular interactions with AlphaFold 3},
  author={Abramson, Josh and Adler, Jonas and Dunger, Jack and Evans, Richard and Green, Tim and Pritzel, Alexander and Ronneberger, Olaf and Willmore, Lindsay and Ballard, Andrew J and Bambrick, Joshua and others},
  journal={Nature},
  volume={630},
  number={8016},
  pages={493--500},
  year={2024},
  publisher={Nature Publishing Group UK London}
}

@article{kim2024enhanced,
  title={Enhanced protein-protein interaction discovery via AlphaFold-Multimer},
  author={Kim, Ah-Ram and Hu, Yanhui and Comjean, Aram and Rodiger, Jonathan and Mohr, Stephanie E and Perrimon, Norbert},
  journal={BioRxiv},
  year={2024}
}

@article{gonzalez2025combinatorial,
  title={Combinatorial prediction of therapeutic perturbations using causally inspired neural networks},
  author={Gonzalez, Guadalupe and Lin, Xiang and Herath, Isuru and Veselkov, Kirill and Bronstein, Michael and Zitnik, Marinka},
  journal={Nature Biomedical Engineering},
  pages={1--18},
  year={2025},
  publisher={Nature Publishing Group UK London}
}
\bibliographystyle{icml2026}

\newpage

\clearpage
\appendix
\setcounter{table}{0}
\renewcommand{\thetable}{S\arabic{table}}
\setcounter{figure}{0}
\renewcommand{\thefigure}{S\arabic{figure}}
\onecolumn
\section{Appendix: Site4Drug Report}
\label{app:reportss}

\begin{figure}[H]
\centering
\begin{subfigure}{\textwidth}
  \centering
  \includegraphics[width=0.9\textwidth]{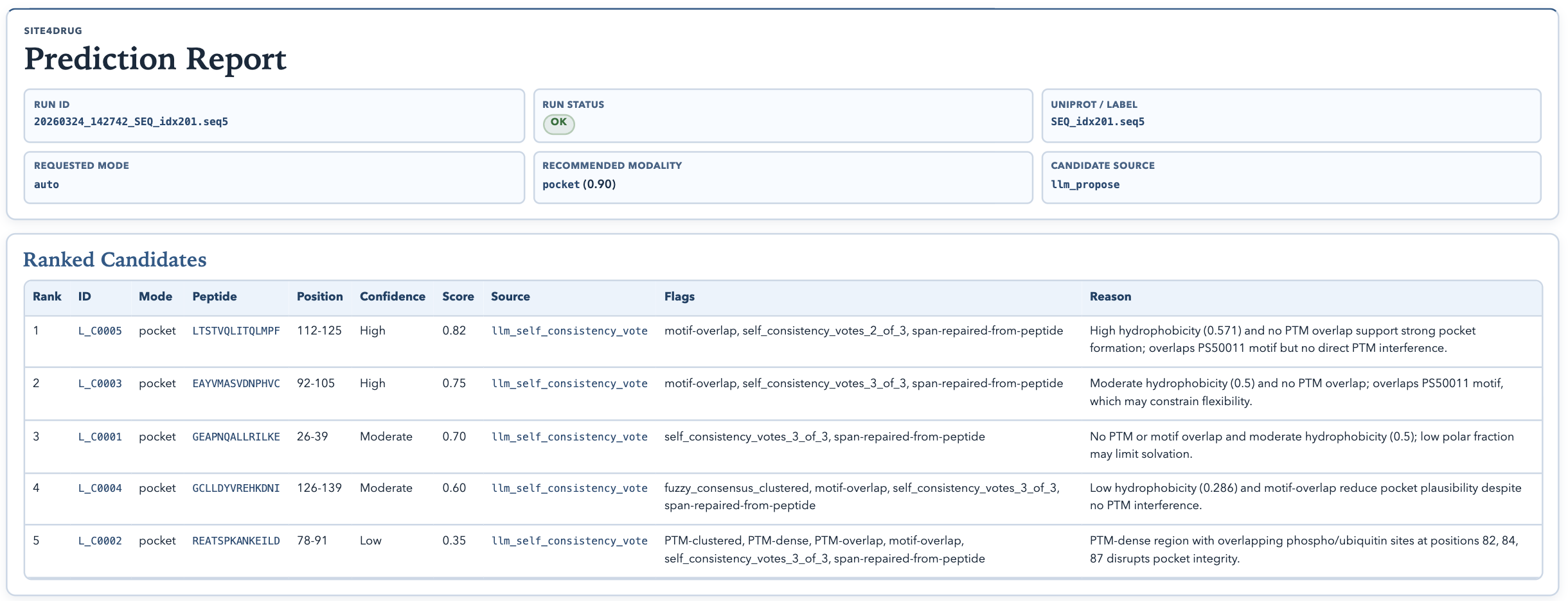}
  \caption{Structured output report with recommended modality, ranked candidates, confidence, flags, and rationale.}
  \label{fig:prediction_report}
\end{subfigure}

\vspace{0.8em}

\begin{subfigure}{\textwidth}
  \centering
  \includegraphics[width=0.9\textwidth]{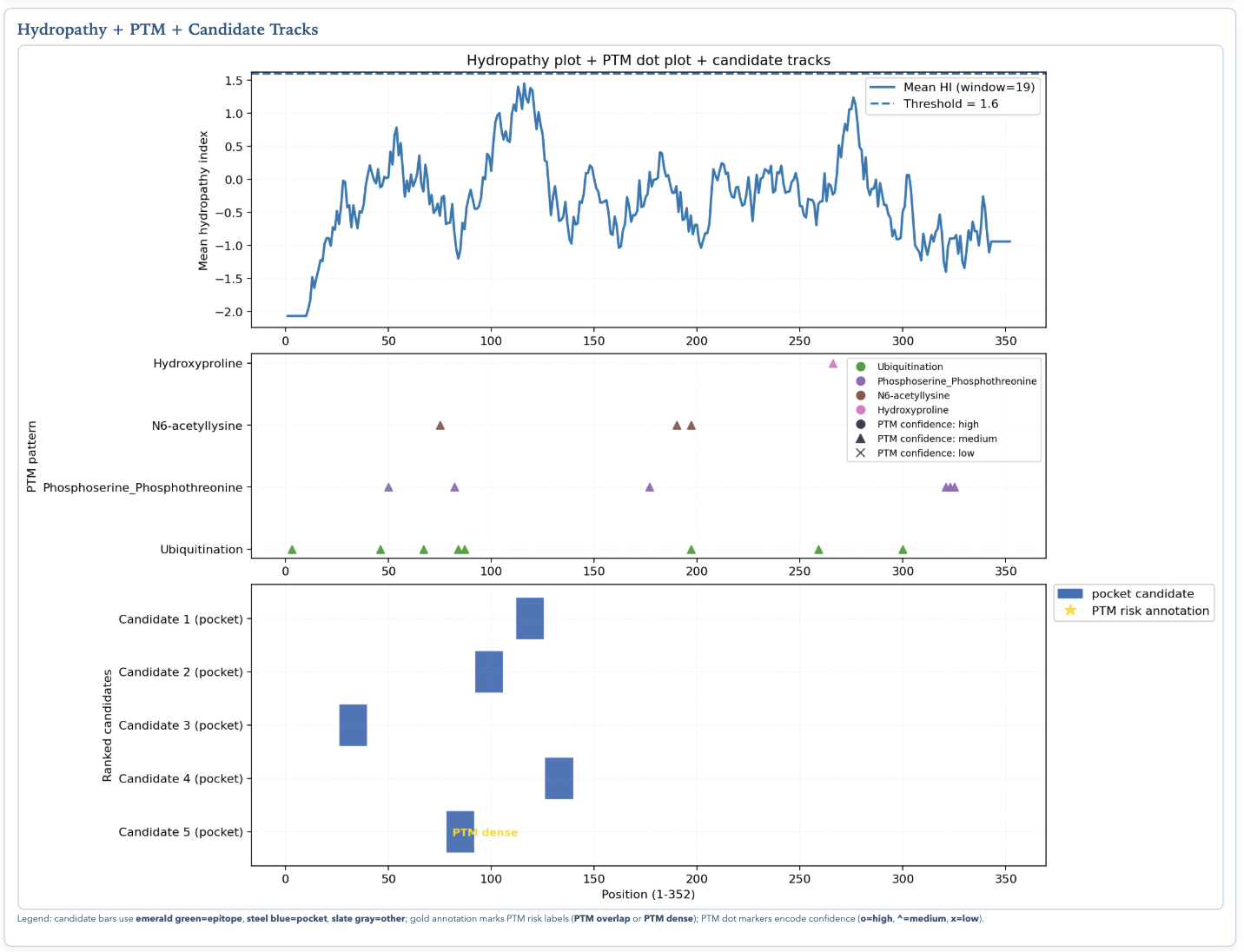}
  \caption{Sequence-level evidence view showing hydropathy, PTM annotations, and candidate tracks.}
  \label{fig:hydropathy_tracks}
\end{subfigure}

\caption{Example \sysname{} prediction interface for Epidermal Growth Factor Receptor (EGFR). (a) Structured output report returned by \sysname{}, including recommended modality, ranked candidates, confidence, flags, and rationale. (b) Corresponding evidence view used to support candidate ranking and risk annotation. Pocket-mode prediction is supported by the observation of lower hydropathy scores. Ground-truth small-molecule binding-site residues are L48, V56, A73, K75, M96, C105, R106, L107, L118, T120, Q121, L122, M123, G126, C127, L129, D130, R133, L174, T184, D185, F186, and L188.}
\label{fig:prediction_interface}

\end{figure}

\section{Appendix: Prompt Templates}
\label{app:prompts}

\paragraph{Inference system prompt.}
\begin{promptbox}
You are Site4Drug, an expert system for constraint-first targetable site discovery on proteins.
You must decide target modality (epitope vs pocket) unless explicitly fixed.
Epitope mode predicts where antibodies or peptide binders attach, usually on extracellular protruding regions of membrane proteins, often tied to immune responses.
Pocket mode predicts where small molecules bind, typically in intracellular proteins or membrane channels, covering immune as well as general signaling and metabolic pathways.
Follow a constraint hierarchy: topology/TM -> PTM masks -> motif-functional caveats -> disulfide/context.
Generate or rank targetable candidates with evidence-grounded rationale.
Always output strict JSON.
\end{promptbox}

\paragraph{Full-sequence inference template.}
Here, \texttt{<target\_id>} denotes a generic target identifier. In benchmark evaluation this field was typically a neutral sequence label (e.g., \texttt{SEQ\_idx001.seq0}).

The default proposal prompt combines the full target sequence with PTM and motif summaries:

\begin{promptbox}
Objective:
Identify targetable site regions using constraint-first analysis and return strict JSON.

Modality policy:
<mode_instruction>

<auto_block_if_mode_is_auto>Constraint hierarchy:
1) TM/topology constraints
2) PTM mask constraints (typed)
3) Motif-functional caveats
4) Disulfide/context risks

Target: UniProt <uniprot>
Length: <len(sequence)> aa
TM regions: <seq_summary.tm_regions>
Cysteines: <len(seq_summary.cysteine_positions)>

<ptm_text>

<motif_text>

Antigen sequence:
<sequence>

Output request:
Return Top-<k> candidates.
<schema_instruction>
\end{promptbox}

If auto mode is enabled, the prompt additionally inserts a deterministic policy block of the form:
\begin{promptbox}
Auto mode policy (deterministic project rule):
- Non-membrane protein -> pocket
- Membrane protein -> pocket or epitope
- This target policy decision: <mode> (n_tm_regions=<n>, channel_like=<bool>)
- Policy reason: <reason>
Set recommended_modality to this policy decision.

When emitting ranked_candidates, prefer row modes that match this policy decision.
\end{promptbox}

\paragraph{Candidate-ranking inference template.}
When candidate preprocessing is available, the system can be configured to ask the model to rank a provided candidate table:
\begin{promptbox}
Objective:
Rank provided candidate regions using constraint-first evidence and return strict JSON.

Modality policy:
<mode_instruction>
- <chunk_note>

<auto_block_if_mode_is_auto>Constraint hierarchy:
1) TM/topology constraints
2) PTM mask constraints (typed)
3) Motif-functional caveats
4) Disulfide/context risks

Target: Sequence ID <target_id>
Length: <seq_summary.sequence_length> aa
TM regions: <seq_summary.tm_regions>
Cysteine count: <len(seq_summary.cysteine_positions)>

<ptm_text>

<motif_text>

Candidate table:
<candidate_table>

Output request:
Return exactly Top-<k>.
<schema_instruction>
\end{promptbox}

\paragraph{Example instantiated evidence snippets.}
The PTM and motif summaries are rendered in the following format:
\begin{promptbox}
<ptm_text> example:
PTM summary: total=3, by_type=N-linked_glycosylation:2, Phosphoserine_Phosphothreonine:1
PTM sites (typed):
- N-linked_glycosylation @ 82 (mask 77-87, confidence=high)
- N-linked_glycosylation @ 144 (mask 139-149, confidence=high)
- Phosphoserine_Phosphothreonine @ 211 (mask 208-214, confidence=medium)

<motif_text> example:
Motif summary: total=2, by_name=EGF_2:1, N-glyco_site:1
Motif hits:
- EGF_2 (PS01186) @ 128-165 [remote_scanprosite_biopython]
- N-glyco_site (PS00001) @ 141-143 [remote_scanprosite_biopython]
\end{promptbox}

\paragraph{Raw proposal JSON schema.}
The top-level proposer is asked to emit a compact JSON object with the following structure:
\begin{promptbox}
{
  "recommended_modality": "epitope|pocket|other",
  "modality_confidence": 0.0,
  "ranked_candidates": [
    {
      "rank": 1,
      "start": 1,
      "end": 15,
      "peptide": "AAAA",
      "mode": "epitope|pocket|other",
      "confidence_score": 0.0,
      "reason": "..."
    }
  ]
}
\end{promptbox}

The schema instruction also requires that ranked candidates remain span-consistent, that PTM/motif caveats be mentioned when present, and that the output begin with \texttt{\{} and end with \texttt{\}} with no extra prose.

\paragraph{Final normalized output artifact.}
After parsing, validation, enrichment, and optional agent-panel adjudication, the run artifact is normalized into a richer schema with the following high-level fields:
\begin{promptbox}
schema_version
recommended_modality
modality_confidence
ranked_candidates[{rank, candidate_id, start, end, peptide, mode, confidence, confidence_score, flags, reason, mean_hydropathy, overlaps_tm, overlaps_ptm_mask, ptm_overlap_by_type, motif_hit_count, cysteine_count}]
candidate_evidence
risk_flags
agent_traces
feature_provenance
token_strategy_used
ptm_summary
motif_summary
orchestrator_trace
audit_log
\end{promptbox}

\paragraph{Specialist-agent critique format.}
Each specialist agent is prompted to return compact JSON of the following form:
\begin{promptbox}
{
  "agent": "...",
  "modality_votes": {"epitope": 0.0, "pocket": 0.0, "other": 0.0},
  "candidate_adjustments": [
    {"candidate_id": "...", "delta": 0.0, "reason": "...", "evidence": ["..."]}
  ],
  "risk_flags": ["..."],
  "summary": "..."
}
\end{promptbox}

\paragraph{Decision-agent format.}
The decision agent uses a different output contract:
\begin{promptbox}
{
  "recommended_modality": "epitope|pocket|other",
  "modality_confidence": 0.0,
  "ranking": [
    {"rank": 1, "candidate_id": "...", "reason": "...", "confidence_score": 0.0, "confidence_reason": "..."}
  ],
  "global_risks": ["..."]
}
\end{promptbox}

\section{Appendix: Hydropathy Profile (Reference Implementation)}
\label{app:hydropathy}
\noindent Reference implementation used by the sequence feature engine to build a center-aligned hydropathy profile.

\begin{promptbox}
HI = {
  "A":1.8, "C":2.5, "D":-3.5, "E":-3.5, "F":2.8,
  "G":-0.4, "H":-3.2, "I":4.5, "K":-3.9, "L":3.8,
  "M":1.9, "N":-3.5, "P":-1.6, "Q":-3.5, "R":-4.5,
  "S":-0.8, "T":-0.7, "V":4.2, "W":-0.9, "Y":-1.3,
}

def sliding_hydropathy(sequence, window=19):
  if not sequence:
    return []
  vals = [HI.get(a, 0.0) for a in sequence]
  if len(vals) < window:
    mean_val = sum(vals) / max(len(vals), 1)
    return [mean_val] * len(vals)

  means = []
  running_sum = sum(vals[:window])
  means.append(running_sum / window)
  for i in range(1, len(vals) - window + 1):
    running_sum += vals[i + window - 1] - vals[i - 1]
    means.append(running_sum / window)

  pad_left = window // 2
  pad_right = len(vals) - len(means) - pad_left
  return [means[0]] * pad_left + means + [means[-1]] * pad_right
\end{promptbox}

\section{Appendix: Sequence-only Ablation}
\label{app:seq_only_ablation}
\noindent Additional prompt and evaluation details for the sequence-only ablation benchmark.

\subsection{Sequence-only Ablation Prompt}
\label{app:seq_only_prompt}
\noindent Exact prompt template used in the sequence-only ablation setting.

To isolate the contribution of the explicit evidence pipeline, we constructed a sequence-only ablation using a prompt that was kept nearly identical to the main full-sequence inference prompt. The same system instruction, \texttt{auto} modality setting, and JSON output contract were retained. The only intentional prompt-level change was to remove the structured evidence fields, including TM/topology summaries, cysteine counts, PTM summaries and typed PTM sites, motif summaries and motif hits, and the deterministic auto-policy block. Here, \texttt{<target\_id>} denotes a generic target identifier; in benchmark evaluation this field was typically a neutral sequence label such as \texttt{SEQ\_idx001.seq0}.

\begin{promptbox}
System:
You are Site4Drug, an expert system for constraint-first targetable site discovery on proteins.
You must decide target modality (epitope vs pocket) unless explicitly fixed.
Epitope mode predicts where antibodies or peptide binders attach, usually on extracellular protruding regions of membrane proteins, often tied to immune responses.
Pocket mode predicts where small molecules bind, typically in intracellular proteins or membrane channels, covering immune as well as general signaling and metabolic pathways.
Follow a constraint hierarchy: topology/TM -> PTM masks -> motif-functional caveats -> disulfide/context.
Generate or rank targetable candidates with evidence-grounded rationale.
Always output strict JSON.

User:
Objective:
Identify targetable site regions using constraint-first analysis and return strict JSON.

Modality policy:
Choose modality automatically. Keep each candidate mode consistent with your final recommended_modality.

Constraint hierarchy:
1) TM/topology constraints
2) PTM mask constraints (typed)
3) Motif-functional caveats
4) Disulfide/context risks

Target: Sequence ID <target_id>
Length: <len(sequence)> aa

Antigen sequence:
<sequence>

Output request:
Return Top-<k> candidates.
<schema_instruction>
\end{promptbox}

\subsection{Sequence-only Ablation Results}
\label{app:seq_only_results}
\noindent Supplementary results for the sequence-only ablation on the included pocket benchmark.
\begin{table}[h]
\centering
\caption{Sequence-only ablation on the included pocket benchmark (\texttt{Group S + Group AS}, $n=63$).}
\label{tab:sequence_only_ablation_appendix}
\setlength{\tabcolsep}{6pt}
\renewcommand{\arraystretch}{1.05}
\begin{tabular}{lccc}
\hline
\textbf{Setting} & \textbf{Predicted pocket} & \textbf{Top-1 sig.} & \textbf{Top-5 sig.} \\
\hline
Sequence-only ($k=1$) & $23/63$ (36.5\%) & $3/63$ (4.8\%) & $3/63$ (4.8\%) \\
Sequence-only ($k=3$) & $22/63$ (34.9\%) & $7/63$ (11.1\%) & $6/63$ (9.5\%) \\
\hline
\end{tabular}
\end{table}

\section{Appendix: Reference Candidate Features and Risk Annotations}
\label{app:feature_ref}

\noindent In the current default \texttt{llm\_propose} inference path, the \LLM\ proposes and ranks candidate spans from the evidence-rich prompt context. The deterministic feature engine is then applied post hoc to each accepted span for candidate enrichment, typed risk annotation, audit logging, constraint checking. The same features are also used directly for ranking.

\paragraph{Candidate-level reference features.}
For a candidate span $r$, we compute mean hydropathy $h(r)$, amino-acid composition fractions, TM overlap, PTM-mask overlap, motif overlap, and cysteine count. Table~\ref{tab:posthoc_features} summarizes the main derived quantities used by the current reference implementation.

\begin{table}[h]
\centering
\small
\caption{Candidate-level deterministic features used for post hoc enrichment and, in deterministic mode, heuristic ranking.}
\label{tab:posthoc_features}
\begin{tabular}{@{}ll@{}}
\toprule
Feature & Definition \\
\midrule
$h(r)$ & mean hydropathy over span $r$ \\
$\phi_{\textsf{hydrophobic}}(r)$ & hydrophobic residue fraction in $r$ \\
$\phi_{\textsf{polar}}(r)$ & polar residue fraction in $r$ \\
$\phi_{+}(r)$ & positively charged residue fraction in $r$ \\
$\phi_{-}(r)$ & negatively charged residue fraction in $r$ \\
$I_{\textsf{TM}}(r)$ & indicator that $r$ overlaps a detected TM segment \\
$c_{\textsf{motif}}(r)$ & number of overlapping PROSITE/ScanProsite motif matches \\
$n_{\textsf{glyco}}(r)$ & number of overlapping N-linked glycosylation masks \\
$n_{\textsf{other}}(r)$ & number of overlapping non-glycosylation PTM masks \\
$d_{\textsf{PTM}}(r)$ & $\bigl(n_{\textsf{glyco}}(r)+n_{\textsf{other}}(r)\bigr)/|r|$ \\
$c_{\textsf{cys}}(r)$ & number of cysteines in $r$ \\
\bottomrule
\end{tabular}
\end{table}

\paragraph{Reference heuristic score.}
Conceptually, these features support a mode-specific preference of the form
\[
S_0(r) = s_{\textsf{mode}}(r) - p_{\textsf{TM}}(r) - p_{\textsf{PTM}}(r) - p_{\textsf{motif}}(r),
\]
although the current implementation uses different exact formulas by mode. Under the default \texttt{tiered} PTM policy, epitope mode favors mildly hydrophilic, polar, non-TM, PTM-light spans, whereas pocket mode favors more hydrophobic spans and penalizes PTM overlap less aggressively.

\paragraph{Typed risk annotations.}
In parallel with scalar score computation, we emit typed risk annotations that preserve \emph{why} a candidate may be unfavorable. These annotations are not treated as interchangeable penalties. Table~\ref{tab:risk_flags_ref} summarizes the current rule triggers.

\begin{table}[h]
\centering
\small
\caption{Reference typed risk annotations and their current trigger conditions.}
\label{tab:risk_flags_ref}
\begin{tabular}{@{}ll@{}}
\toprule
Risk flag & Trigger condition \\
\midrule
\texttt{TM-overlap} & $I_{\textsf{TM}}(r)=1$ \\
\texttt{PTM-overlap} & $n_{\textsf{glyco}}(r)+n_{\textsf{other}}(r) > 0$ \\
\texttt{glyco-mask-overlap} & $n_{\textsf{glyco}}(r) > 0$ \\
\texttt{PTM-dense} & $d_{\textsf{PTM}}(r)\ge 0.25$ or total overlapping PTMs $\ge 3$ \\
\texttt{PTM-clustered} & $d_{\textsf{PTM}}(r)\ge 0.25$ or total overlapping PTMs $\ge 3$ \\
\texttt{disulfide-constrained} & $c_{\textsf{cys}}(r)\ge 2$ \\
\texttt{hydrophobic-core} & $h(r) > 1.0$ \\
\texttt{motif-overlap} & $c_{\textsf{motif}}(r) > 0$ \\
\bottomrule
\end{tabular}
\end{table}

\noindent Additional motif-specific flags may be emitted when motif names contain particular substrings; for example, zinc- or DNA-related motif names produce extra cautionary annotations in epitope mode. Because explicit disulfide bond pairing is not modeled, \texttt{disulfide-constrained} should be interpreted as a cysteine-based structural-risk proxy rather than a direct bond-level call.

\section{Appendix: Illustrative Behavior Shift After Post-Training}
\label{app:format_shift}

\paragraph{Quantitative summary.}
For this target, both models returned 10 ranked candidates. In the raw top-level output, the mean candidate-level reason length decreased from 20.6 words in the base model to 5.5 words in the SFT checkpoint. It gives a more concise formatted "reason" for each candidate.

\paragraph{Original base-model raw output (abridged).}
\begin{promptbox}
"recommended_modality": "pocket",
"modality_confidence": 1.0,
"ranked_candidates": [
  {
    "rank": 1,
    "start": 619,
    "end": 633,
    "peptide": "TPHIDNLEVANSFAV",
    "mode": "pocket",
    "confidence_score": 0.91,
    "reason": "High-confidence phosphosite cluster at 624 and 628; no TM or motif conflicts; region is solvent-accessible and flexible, favorable for small molecule binding; PTM mask 619-633 reduces risk of interference"
  },
  {
    "rank": 2,
    "start": 724,
    "end": 738,
    "peptide": "DYDDSLVSKFIERWS",
    "mode": "pocket",
    "confidence_score": 0.89,
    "reason": "Dense phosphorylation cluster (729, 733, 737) with medium confidence; part of regulatory region; PTM mask 724-738 suggests dynamic regulation; suitable for allosteric pocket targeting"
  },
  ...
]
\end{promptbox}

\paragraph{SFT checkpoint raw output (abridged).}
\begin{promptbox}
{
  "recommended_modality": "pocket",
  "modality_confidence": 0.60,
  "ranked_candidates": [
    {
      "rank": 1,
      "start": 1,
      "end": 15,
      "peptide": "NSIQIGGLFPRGADQ",
      "mode": "epitope",
      "confidence_score": 0.40,
      "reason": "PTM-overlap (Phosphoserine_Phosphothreonine at 2, Methylarginine at 59)"
    },
    {
      "rank": 2,
      "start": 16,
      "end": 30,
      "peptide": "EYSAFRVGMVQFSTS",
      "mode": "epitope",
      "confidence_score": 0.40,
      "reason": "PTM-overlap (Phosphoserine_Phosphothreonine at 29)"
    },
    ...
  ]
}
\end{promptbox}

\paragraph{Original base-model final returned output (abridged).}
\begin{promptbox}
Rank 1 | pocket | TPHIDNLEVANSFAV | 35--49
Reason: High hydrophobic fraction (0.533) and no PTM overlap support strong pocket binding potential.

Rank 2 | pocket | QFGGAEVSGFQIVDY | 227--241
Reason: Favorable mean hydropathy (0.2) and no PTM interference enhance pocket suitability.
\end{promptbox}

\paragraph{SFT checkpoint final returned output (abridged).}
\begin{promptbox}
Rank 1 | epitope | NSIQIGGLFPRGADQ | 1--15
Reason: PTM-overlap (Phosphoserine\_Phosphothreonine at 2, Methylarginine at 59).

Rank 2 | epitope | EYSAFRVGMVQFSTS | 16--30
Reason: PTM-overlap (Phosphoserine\_Phosphothreonine at 29).
\end{promptbox}

\clearpage
\section{Appendix: Module 2 DrugCLIP Results}

\begin{table}[htbp]
\centering
\scriptsize
\caption{Top-6 DrugCLIP hits represented as SMILES strings.}
\begin{tabular}{@{}ll@{}}
\toprule
Rank & SMILES \\
\midrule
Rank-1 & \texttt{CCCN(C)c1nc(SCc2ccc(C(=O)N3CCN(C(C)=O)CC3)cc2)nc(C)c1C} \\
Rank-2 & \texttt{CCCc1c(C\#N)c(N)nc(SCCCN2CCN(c3cccc(Cl)c3)CC2)c1C\#N} \\
Rank-3 & \texttt{CCCCNC(=O)CSc1cc(N2CCN(c3ccccc3OC)CC2)ncn1} \\
Rank-4 & \texttt{COc1ccc(N2CCN(c3cc(SCC(=O)NCCC(C)C)ncn3)CC2)cc1} \\
Rank-5 & \texttt{Cc1sc2nc(CCc3ccccc3)nc(N3CCN(C(=O)Nc4cccc(C\#N)c4)CC3)c2c1C} \\
Rank-6 & \texttt{CCCCc1ccc(NC(=O)CSc2cc(N3CCN(c4ccccc4OC)CC3)ncn2)cc1} \\
\bottomrule
\end{tabular}
\label{tab:top6_smiles}
\end{table}

\begin{figure}[htbp]
  \centering
  \includegraphics[width=0.95\textwidth]{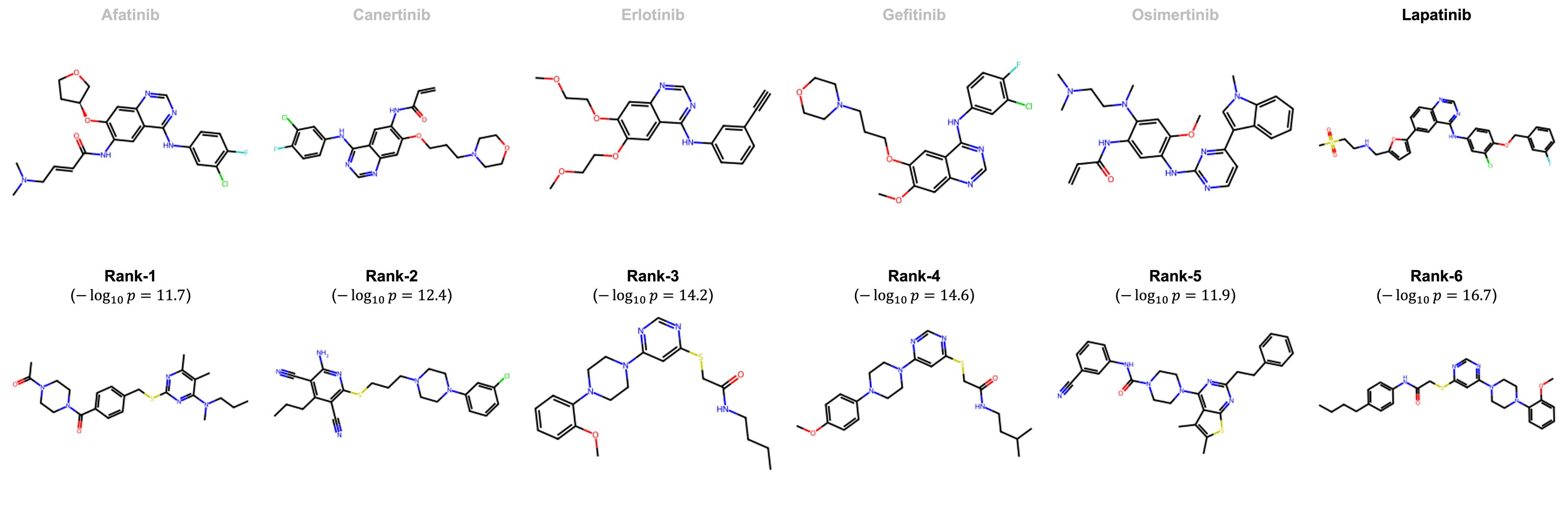}
  \caption{
  Comparison of reference EGFR inhibitors and the top-6 compounds retrieved by DrugCLIP for the EGFR pocket.
  The top row displays the structures of afatinib, canertinib, erlotinib, gefitinib, osimertinib, and lapatinib, and the bottom row displays the top-6 DrugCLIP hits. The input FASTA sequence was derived from RCSB 1XKK (EGFR bound to lapatinib). Despite the absence of the known EGFR inhibitors from the DrugCLIP ligand database, the top-ranked compounds exhibit structurally related chemical motifs, indicating that DrugCLIP captures pocket-compatible structural patterns. Figures were drawn by rdkit. 
  }
  \label{fig:drugclip_egfr}
\end{figure}

\section{Appendix: Module 2 BoltzGen Results}
\begin{figure}[htbp]
  \centering
  \includegraphics[width=0.98\textwidth]{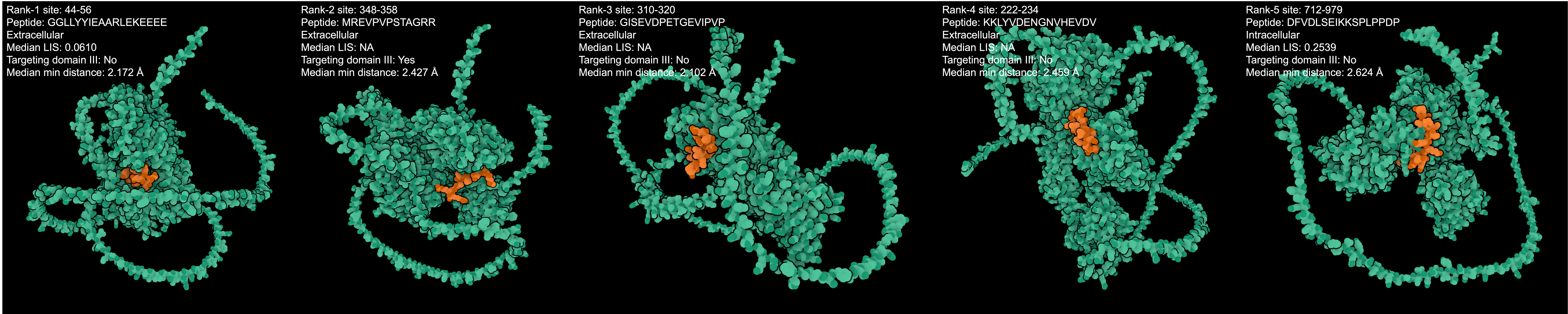}
  \caption{Peptide binder results generated from the top-5 epitopes recommended by \sysname{} in the EGFR precursor. EGFR is initially translated as a 1,210-aa precursor, after which an approximately 24-aa N-terminal signal peptide is removed to produce the 1,186-aa mature protein. Based on the mature protein sequence, the extracellular region consists of domain I (1-165 aa), domain II (166-310 aa), domain III (311-480 aa), and domain IV (481-620 aa), while the intracellular region begins at residue 643. The sites recommended by \sysname{} are consistent with the ranking of the final binder epitopes as assessed by the membrane topology of epitope, the median LIS, the domain III relatedness, and the median minimum distance, in order. EGFR and peptide binders are shown in green and orange. }
  \label{fig:boltzgen}
\end{figure}

\newpage
\section{Appendix: Integration with Diffuse Bio Sandbox}

\begin{figure}[htbp]
    \centering
    \begin{subfigure}[t]{0.58\textwidth}
        \centering
        \includegraphics[width=\linewidth,height=0.20\textheight,keepaspectratio]{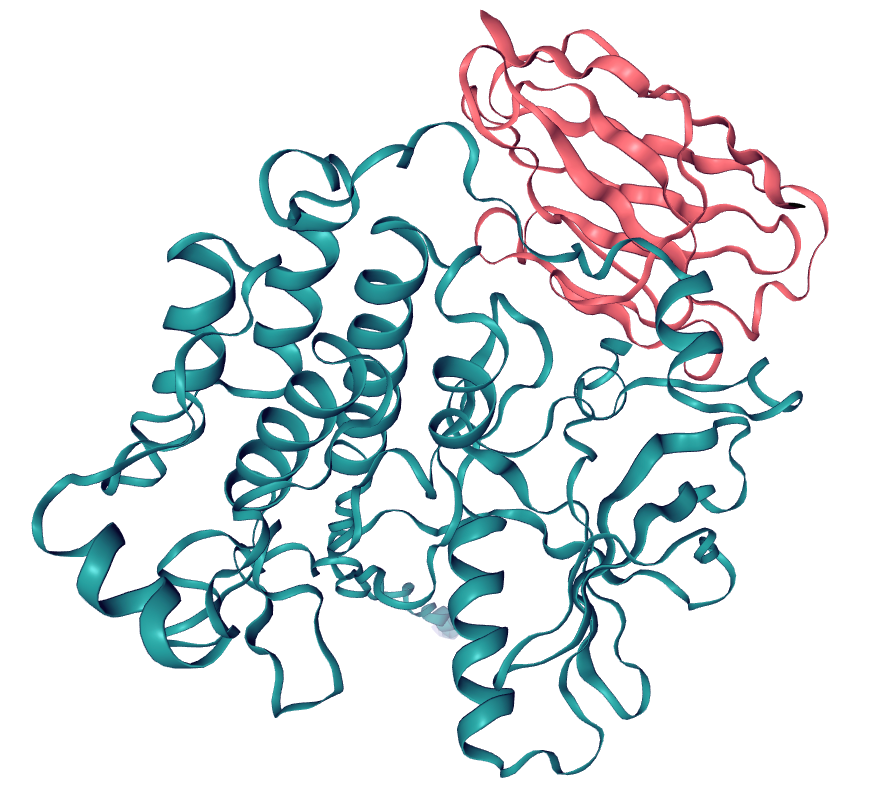}
        \caption{Nanobody binder designed in Diffuse Bio Sandbox.}
        \label{fig:egfr_nanobody}
    \end{subfigure}
    \hfill
    \begin{subfigure}[t]{0.38\textwidth}
        \centering
        \includegraphics[width=\linewidth,height=0.2\textheight,keepaspectratio]{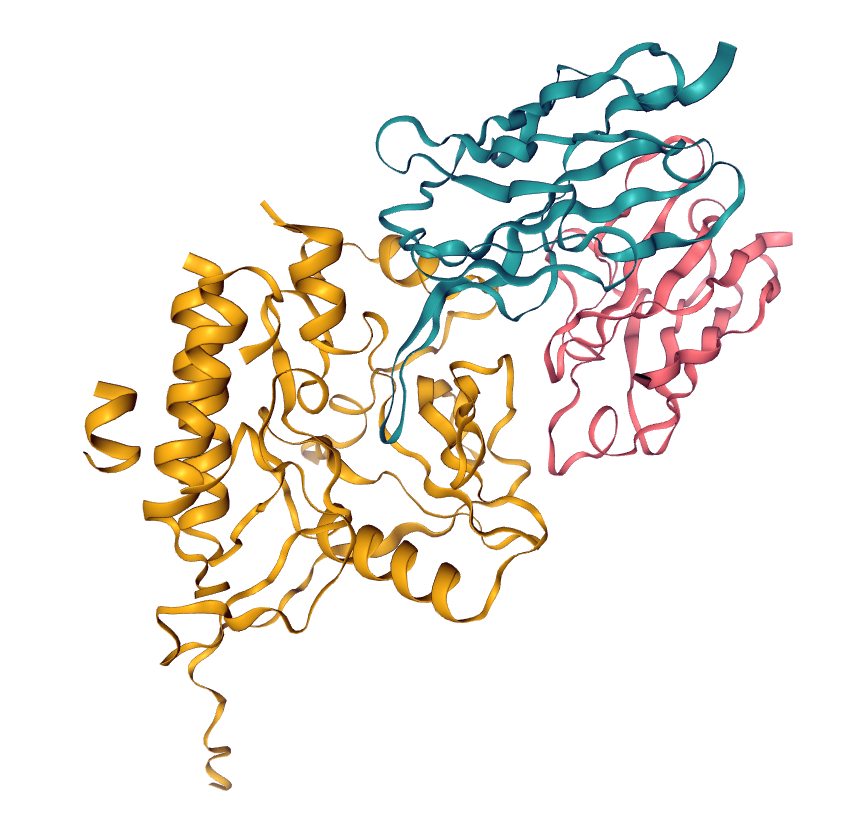}
        \caption{scFv binder designed in Diffuse Bio Sandbox.}
        \label{fig:egfr_scfv}
    \end{subfigure}

    \caption{
    Integration with Diffuse Bio Sandbox for structure-conditioned binder generation using the \sysname{} epitope candidate on EGFR (RCSB code: 1XKK). 
    The selected epitope residues are 34--36, 41, 55--60, 70--75, 92, 105--110, and 115--121, which are aligned with top-5 sites by \sysname{}: 112-125, 92-105, 26-39, 126-139, and 78-91. 
    (a) Nanobody design generated against the specified epitope. Green and red parts correspond to EGFR and designed nanobody drug, respectively. (b) scFv design generated against the same epitope. Yellow and the other parts correspond to EGFR protein and designed scFv drug, respectively. 
    These examples illustrate how Site4Drug-selected target regions can be passed downstream to external binder-generation frameworks. Figures were generated by Diffuse Bio Sandbox visualization tool.
    }
    \label{fig:egfr_diffusebio_appendix}
\end{figure}

\clearpage
\section{Appendix: Audit Logs as a Tool for Identifying Binding Failure Modes}
\label{app:failuremode}

The following example summarizes how the \sysname{} audit log can support interpretation of a binding failure mode.

\begin{promptbox}

Rank 1 | epitope | candidate region

Reason: The candidate was ranked first because it showed no PTM mask overlap and low PTM density, suggesting a PTM-free target region with high initial confidence. However, the region was also highly hydrophobic, with mean_hydropathy = 0.9733 and hydrophobic_fraction = 0.6. ChemAgent flagged this high hydrophobicity as a potential risk because it may reduce solubility and epitope accessibility.

Interpretation: Although the predicted epitope appears favorable from a PTM perspective, its high hydrophobicity may indicate limited exposure in the native protein or cell-surface context. Therefore, the candidate may be detectable as a linear sequence-based epitope but less accessible in the folded protein.

Mechanistic hypothesis: After antibody mutation, a subtle change in paratope geometry may have reduced tolerance to this hydrophobic and partially exposed epitope. The WT antibody may still recognize the region weakly, whereas the mutant antibody may no longer bind it stably.

Additional context: The candidate may not function as an isolated linear epitope, but rather as part of a conformational epitope shaped by neighboring structural features. Full-length sequence analysis identified multiple risk flags, including PTM-dense regions, glyco-mask overlap, and disulfide-constrained regions, suggesting that folding, glycosylation, and structural context may influence binding.

Conclusion: The binding loss after antibody mutation is therefore more consistent with reduced recognition of a partially exposed or structure-dependent epitope than with an incorrect epitope sequence prediction.

Suggested follow-up: Compare WT and mutant antibodies using peptide ELISA, recombinant extracellular domain binding, and cell-based binding assays. Binding in peptide ELISA but not in full-length or cell-based assays would support an accessibility or conformational epitope issue. Loss of binding even in peptide ELISA would instead suggest that the paratope mutation disrupted affinity for the linear epitope itself.
\end{promptbox}

\end{document}